\documentclass[11pt,a4paper]{article}
\usepackage{jheppub}

\usepackage{mathrsfs}
\usepackage{url}
 
\usepackage{amsthm}

\newcommand{\integers}{\ensuremath{\mathbb{Z}}}

\newcommand{\real}{\ensuremath{\mathbb{R}}}

\newcommand{\bigo}{\ensuremath{\mathcal{O}}}

\newcommand{\vect}[1]{\ensuremath{\boldsymbol{#1}}}
\newcommand{\barr}[1]{\ensuremath{\overline{#1}}}

\newcommand{\eqdef}{\ensuremath{:=}}
\newcommand{\defeq}{\ensuremath{=:}}

\newcommand{\extder}{\ensuremath{\mathbf{d}}}
\newcommand{\weq}{\ensuremath{\approx}}
\newcommand{\lie}{\ensuremath{\mathscr{L}}}
\newcommand{\liephase}{\ensuremath{\boldsymbol{\mathcal{L}}}}

\newcommand{\Poi}{\mathsf{Poi}}
\newcommand{\SU}{\mathrm{SU}}
\newcommand{\U}{\mathrm{U}}

\newcommand{\realpart}{\ensuremath{\textrm{Re}}}
\newcommand{\imaginarypart}{\ensuremath{\textrm{Im}}}
\begin{document}
\title{Asymptotic symmetries of scalar electrodynamics and of the abelian Higgs model in Hamiltonian formulation}
\author[a]{Roberto Tanzi}%\footnote{Corresponding author: v}}
\author[a,b]{and Domenico Giulini}
\affiliation[a]{University of Bremen, Center of Applied Space Technology and Microgravity (ZARM), 28359 Bremen}
\affiliation[b]{Leibniz University of Hannover, Institute for Theoretical Physics, 30167 Hannover, Germany}
\emailAdd{roberto.tanzi@zarm.uni-bremen.de}
\emailAdd{giulini@itp.uni-hannover.de}
\arxivnumber{2101.07234}

%\dedicated{}

\abstract{
 We investigate the asymptotic symmetry group of a scalar
field minimally-coupled to an abelian gauge field using 
the Hamiltonian formulation. This extends previous work 
by Henneaux and Troessaert on the pure electromagnetic 
case. We deal with minimally coupled massive and
massless scalar fields and find that they behave
differently insofar as the latter do not allow for
canonically implemented asymptotic boost symmetries.  
We also consider the abelian Higgs model and show that 
its asymptotic canonical symmetries reduce to the
Poincar\'e group in an unproblematic fashion.}

\maketitle

% \newpage
% \tableofcontents
% \newpage

\section{Introduction} \label{sec:introduction}
With this paper, we wish to continue our previous work
\cite{Tanzi:2020} on the asymptotic structure and 
symmetries of Yang-Mills gauge theories, which was 
motivated by the seminal work of Henneaux and 
Troessaert on various field theories, including the 
pure electromagnetic case. We refer to 
\cite{Tanzi:2020} for a more extended list of references. In our previous paper, we investigated 
the case
of pure $\SU(N)$-Yang-Mills-theory and found that globally-charged states cannot exist in that theory if 
its phase space is to be endowed with well-defined 
symplectic structure, a well-defined Hamiltonian 
vector field driving its evolution, and a well-defined canonical \mbox{(Poisson-)} action of the whole 
identity-component of the Poincar\'e group. When we
established this result, which we initially did not 
expect, we welcomed it as a potential classical analogue
(or sign) of confinement. At the same time, we asked
ourselves whether one must fear that this exclusion 
of globally-charged states may also occur in  
other field theories and in situations in which 
physics actually requires such states to exist.
Clearly, if that turns out to happen it would 
cast serious doubts on the method we employed. 
Hence, we set ourselves the goal to check other 
field theories in order to see --- hopefully --- 
how the Hamiltonian formalism leads to results 
compatible with these expectations.

Our overall plan is to start with simple models,
gradually including more fields of physical significance. In doing that, we follow the 
Hamiltonian strategy pioneered by Henneaux and Troessaert, who already made a detailed 
investigation into the electromagnetic case~\cite{Henneaux-ED} (even in 
higher dimensions~\cite{Henneaux-ED-higher}). 
As an obvious generalisation of their work 
we decided to look at the case of electromagnetism 
coupled to a scalar field. This is what the 
present paper is about.

More precisely, we deal with two main cases, with 
two subcases in the first. In the first main case, 
we consider what is commonly referred to as 
``scalar electrodynamics''. That is,  a scalar 
field endowed with a potential which, depending 
on its precise form, represents either a 
massless (first subcase) or a massive 
(second subcase) scalar field, minimally-coupled 
to the electromagnetic fields. Interestingly, 
the outcome of our analysis crucially depends 
on whether or not the scalar field  has a mass. 
We show that a massive field has to decay 
at infinity faster than any power-like function
in the affine coordinates, so that the behaviour
of the electromagnetic fields, as well as the 
symmetry group, is the same as the one found by 
Henneaux and Troessaert in the case of free 
electrodynamics \cite{Henneaux-ED}.
On the other hand, a massless scalar field renders 
the boosts of the Poincar\'e transformations 
non canonical in a way which is difficult to 
circumvent, leading either to a trivial 
asymptotic symmetry group or to a non-canonical 
action of the  Poincar\'e group. We highlight 
a connection of this problem with the 
impossibility of a Lorenz gauge-fixing if the 
flux of  charge-current at null infinity is 
present, as  pointed out by Satishchandran and Wald \cite{Wald-Satishchandran}. All this is derived in 
Section\,\ref{sec:scalar-electrodynamics}.

As our second main case, we consider the abelian Higgs
model, i.e., a potential of the scalar field which 
leads to spontaneous symmetry breaking, thereby 
reducing the $\U(1)$ gauge-symmetry group to 
 the trivial group. We show that the asymptotic 
symmetry group  reduces in a straightforward way 
to the Poincar\'e  transformations without any complications. All this is derived in 
Section\,\ref{sec:abelian-Higgs}. 

Section \ref{sec:Lagrangian-Hamiltonian} sets up 
the Hamiltonian formalism for the present context 
and shows how to canonically implement the 
Poincar\'e action. Section\,\ref{sec:free-scalar}
introduces the scalar-field models with a brief 
digression of the free scalar field for illustrative
purposes. The final Section\,\ref{sec:conclusions}
concludes and also contains some speculations on 
what might happen in the physically relevant 
electroweak case.  Appendix\,\ref{appendix:massive-fall-off} contains the proof of the statement that in the 
massive case the scalar field as well as its 
momentum fall-off faster than any power in the 
affine coordinates.

\subsection*{Conventions and notation}
\noindent
Throughout this paper, we adopt the following conventions.
Lower-case Greek indices denote spacetime components, 
e.g. $\alpha=0,1,2,3$, lower-case Latin indices denote 
spatial  components, e.g. $a=1,2,3$, and lower-case 
barred Latin indices denote angular components, e.g. 
$\bar a=\theta,\varphi$. 
We adopt the mostly-plus convention $(-,+,+,+)$ for 
the spacetime four-metric ${}^4 g$.

\section{Hamiltonian and Poincar\'e transformations} 
\label{sec:Lagrangian-Hamiltonian}
In this section, we provide the basic tools that are necessary for the Hamiltonian treatment of an abelian gauge field minimally-coupled to a scalar field.
Specifically, we are going to identify the canonical fields and momenta, derive the Hamiltonian and the symplectic form, and infer the action of the Poincar\'e transformations on the canonical fields.
We will postpone to the next sections the discussion concerning whether or not the derived quantity are well defined, which usually amounts to the correct identification of the phase space --- by imposing fall-off and parity conditions --- and to the completion of the Hamiltonian generators and of the symplectic form by adequate surface terms.
For now, we will assume that these quantities are well defined, in order to allow the following formal manipulations.

We start our discussion from the more-familiar Lagrangian picture, in which the action reads
\begin{equation} \label{action-Lagrangian}
\begin{aligned}
 S[A_\alpha,\dot A_\alpha, \varphi, \dot \varphi;g] ={}&
  \int d^4 x \sqrt{-{}^4g} \left[
  -\frac{1}{4} {}^4g^{\alpha \gamma} \, {}^4g^{\beta \delta} \, F_{\alpha \beta} F_{\gamma \delta}
 -{}^4g^{\alpha \beta} \big( D_{\alpha} \varphi \big)^* D_{\beta} \varphi
 -V(\varphi^* \varphi)
 \right] + \\
 &+(\text{boundary terms})\,,
\end{aligned}
\end{equation}
where $A$ is the one-form abelian potential,  $F \eqdef d A$ is the curvature (or field strength) two-form, $\varphi$ is a complex scalar field, and ${}^4 g$ is the four-dimensional flat spacetime metric.
Moreover,
\begin{equation}
 D_\alpha \varphi \eqdef \partial_\alpha \varphi + i A_\alpha \varphi
\end{equation}
is the gauge-covariant derivative in the fundamental representation and the potential $V(\varphi^* \varphi)$ is explicitly given by the expression
\begin{equation} \label{potential}
 V(\varphi^* \varphi) \eqdef -\mu^2 \varphi^* \varphi +\lambda (\varphi^* \varphi)^2 \,,
\end{equation}
where $\lambda$ and $\mu^2$ are two real parameters.
In this paper, we wish to analyse two specific situations, which arise depending on the value of these two parameters.

The former situation is \emph{scalar electrodynamics}.
Namely, it corresponds to the case in which the two parameters appearing in the potential~(\ref{potential}) are such that $m^2 \eqdef -\mu^2 \ge 0$ and $\lambda \ge 0$.
As we shall see in section~\ref{sec:scalar-electrodynamics}, there are going to be some important differences in the asymptotic structure of the theory depending on whether we are dealing with a massless scalar field ($m^2 = 0$) or with a massive one ($m^2 > 0$).
Note that we are leaving open the possibility of having a self-interaction term for the scalar field ($\lambda > 0$), as this does not affect our analysis of the asymptotic structure.

The latter situation that we wish to analyse is the \emph{abelian Higgs model}.
This corresponds specifically to the case in which the two parameters appearing in the potential~(\ref{potential}) are such that $\mu^2 > 0$ and $\lambda >0$.
This choice leads to the well-known Mexican-hat shape of the potential and, ultimately, to the spontaneous symmetry breaking of the $\U(1)$ gauge symmetry.\footnote{
Note that the case $\lambda < 0$ needs to be excluded on physical grounds, as it would lead to a Hamiltonian which is not bounded from below.
For the same reason, we have to exclude the case $\lambda = 0$ when $\mu^2 > 0$, which is precisely the setup used in the abelian Higgs model.
}

Finally, let us point out that we have included an undetermined boundary term in the action~(\ref{action-Lagrangian}), which ought to be chosen such that the variation principle is well defined.
For now, we merely assume that a boundary leading to a well-defined action principle exists and postpone to the next sections a thorough discussion about whether or not this assumption is in fact correct.

\subsection{(3+1) decomposition}

Here, we briefly recall the general geometric idea 
that underlies the Hamiltonian formulation of 
field theory. A very detailed exposition may be 
found in \cite{Kuchar:1976c} relying on  \cite{Kuchar:1976a,Kuchar:1976b}. A shorter 
summary is in \cite{Giulini:SpringerHandbookSpacetime}.
Suppose we are given a four-dimensional 
spacetime $(M,{}^4 g)$, that is a four-dimensional 
manifold $M$ with Lorentzian metric ${}^4 g$. Let 
$\Sigma$ be a three-dimensional manifold and 
$E_t:\Sigma\hookrightarrow M$ a one-parameter 
family of spacelike (with respect to ${}^4 g$) 
embeddings of $\Sigma$ into $M$. The parameter 
is $t\in I\subseteq\mathbb{R}$ where $I$ is an open interval. It plays the role of a mathematical  
evolution parameter whose relation with any 
physically meaningful concept of time may depend 
on the circumstances. If, for example, the images 
$\Sigma_t=E_t(\Sigma)$ are such that for a 
fixed point $p\in\Sigma$ the worldline 
$t\mapsto E_t(p)$ is timelike, then an increment 
$dt$ is related to proper time along this 
worldline by a function called the `lapse function'
$N$. Any configuration of covariant fields on 
$M$, like the metric ${}^4g_{\alpha\beta}$, 
the electromagnetic potential $A_\alpha$, or 
the electromagnetic Faraday tensor $F_{\alpha\beta}$ 
of field strengths, can at each value of $t$ 
be pulled back via $E_t^*$ to the corresponding 
quantities on $\Sigma$, which will then be 
collections of tensors of the same and lower 
ranks. For example, the 10 components of the 
four-dimensional Lorentzian metic 
${}^4g_{\alpha\beta}$ will be pulled back to 
the 6 components $g_{ab}$ of a Riemannian 
metric on $\Sigma$ (corresponding to the spatial
components of ${}^4g$), the 3 components of 
a one-form $N_a$ on $\Sigma$ called `shift' (corresponding to the mixed space-time components of ${}^4g$) and a scalar field $N$ 
called `lapse' (related to the time-time 
components of ${}^4g$; see formula \eqref{4-metric}
below). Likewise the four-dimensional 
vector potential $A_\alpha$ is pulled back to a 
one-form $A_a$ on $\Sigma$ (corresponding to the 
spatial components of $A_{\alpha}$) and a scalar
field $A_0$ (corresponding to the time component 
of $A_{\alpha}$). Finally, the Faraday tensor $F_{\alpha\beta}$ will be pulled back to the magnetic 
2-form  $F_{ab}$, or, equivalently in 3-dimensions
due to Hodge duality, a magnetic one form $B_a$, 
and the electric one-form $E_a$ (corresponding to 
the mixed space-time components of $F_{\alpha\beta}$). 

Note that due to the existence of the Lorentzian 
metric ${}^4g$ on $M$, \emph{any} tensor on $M$ 
can be pulled back by first turning it to a purely 
covariant one using ${}^4g_{\alpha\beta}$ (all 
indices pulled down). Moreover, due to the 
Riemannian metric $g_{ab}$ on $\Sigma$, a 
covariant (i.e. those tensor arriving at $\Sigma$ 
via pull back) tensor on $\Sigma$ 
can be turned into contravariant or mixed one 
by raising indices using $g^{ab}$. In this 
fashion, a single rank-$n$ tensor field on $M$ 
at some instant of parameter value $t$ (the 
`instant' is represented in spacetime by the 
spacelike hypersurface $E_t(\Sigma)\subset M$) 
will be faithfully represented by the collection 
of tensors fields on $\Sigma$ of rank $n$ 
and less, that result via the procedure just 
outlined. 

If we let the parameter $t$ run trough 
an open intervall $I\subset\mathbb{R}$ corresponding to 
different embeddings $E_t:\Sigma\hookrightarrow M$, 
we obtain a one-parameter family of tensor
fields on $\Sigma$ corresponding to the 
pull-backs for each $t$. If the system of tensor 
fields on $M$ obeys differential equations 
allowing for a well-posed Cauchy initial-data formulation, these differential equations will
translate into Hamiltonian equations of motion
for the pulled-back fields on $\Sigma$.  The idea
of Hamiltonian field theory is to turn this around 
and to deduce solutions to the hyperbolic equations 
on $M$ from solutions to the Hamiltonian evolution equations on $\Sigma$. Hence, in the Hamiltonian
setting, there is no spacetime to start with, but 
only a three-dimensional Riemannian manifold with
a collection of tensor fields depending on some 
parameter $t$ with respect to which the fields 
obey a set of Hamiltonian equations of motion.  
Once these equations are solved, we can construct a 
four-manifold $M=I\times\Sigma$ and tensor-fields
on it that satisfy the hyperbolic equations. 
For this to produce a finite-time evolution 
everywhere in space, the lapse function should 
be non-zero everywhere. In that case the Lorentz 
metric will assume the form 
\begin{equation} 
\label{4-metric}
 {}^4g_{\alpha \beta}=
 \left(
 \begin{array}{c|c}
  -N^2+g^{ij} N_i N_j	& N_b	\\ \hline
  N_a	& g_{ab}
 \end{array}
 \right) \,.
\end{equation}
which is a Lorentz metric on $I\times\Sigma$
only if $-N^2+g^{ij} N_i N_j<0$. 

For our application it is important to realise 
that the Hamiltonian formalism is not only suited 
to set up the Cauchy problem and find solutions
to it, but also to investigate existing solutions 
in terms of various one-parameter embeddings 
$E_t$ and its associated Hamiltonian generators.
In that case the parameter $t$ labels the different 
embeddings $E_t$ just as before, but the embeddings are not 
restricted to those where images 
$\Sigma_t:=E_t(\Sigma)$ are nicely stacked on top
of each other, so as to really represent a 
nowhere vanishing advancement in physical time. 
We may, for example, consider 
families of embeddings which differ only by a 
rotation of the manifold $\Sigma$, i.e. 
$E_t=E_*\circ R_t$, where $E_*$ is a fixed 
embedding and $R_t$ is an orthogonal rotation 
on $\Sigma$ (i.e. if $\Sigma$ is identified
with euclidean $\mathbb{R}^3$, as well shall do 
later on). Or $\Sigma_t$
may correspond to a family of embeddings in which 
the images $\Sigma_t$ are tilted relative to 
each other and therefore all intersect, like for
boost transformations. We shall continue to label 
the parameter by $t$ in these cases, keeping in mind 
that the possibility and, as the case may be,  quantitative expression of its relation to a 
meaningful physical time parameter depends on the context. 

In our investigation the significance of the 
parameter $t$ is that it appears as the label that
parametrises integral curves of Hamiltonian 
vector fields in phase space, along which we
compute, e.g., the action associated to the  
change of states along the curve. The Hamiltonian 
generating this motion in $t$ will depend on
four functions: one for the lapse $N$ and three 
for the shift $\vect{N}$. Certainly, if we were interested only in finding the time evolution of the fields, it would be possible --- and simpler --- 
to just fix $N=1$ and $\vect{N} = 0$.
In this case, the Dirac procedure~\cite{Dirac-book} would lead us from the action~(\ref{action-Lagrangian}) to the Hamiltonian $H$ generating proper-time 
evolution of the fields through the 
Poisson brackets. But we are interested to not 
only derive that type of time evolution. We also 
wish to determine the transformation properties 
of the fields under the Poincar\'e  transformations,
and this can be done if other, more general choices 
for lapse and shift are made, though we do not 
need, and will not further consider, the most 
general choices. In fact, throughout the paper 
we shall restrict spacetime $(M,{}^4 g)$ to be Minkowski space, $\Sigma$ 
to be diffeomorphic to $\mathbb{R}^3$, and the 
embeddings $E_t$ to be such that each 
image $\Sigma_t:=E_t(\Sigma)\subset M$ is a 
flat spacelike plane. For each parameter $t$ 
the pull-back metric $g_{ab}$ on $\Sigma$ will 
then be be flat. 

The same procedure that let us derive the Hamiltonian
$H$ from the action~(\ref{action-Lagrangian}) 
now produces a Hamiltonian function 
$H[N,\vect{N}]$ generating the transformations on the 
fields via its Hamiltonian flow, that is,
via Poisson bracket. For simplicity we will refer 
to  $H[N, \vect{H}]$ as the generator of 
``hypersurface deformations'', abbreviating the more
cumbersome ``generator of transformations of the
fields under hypersurface deformations''. 

A generating vector field $\xi$ for Poincar\'e 
transformations in Minkowski space may be 
decomposed with respect to a spatial hyperplane 
into its components perpendicular and parallel 
to it. This corresponds to a decomposition 
of lapse and shift 
$(N=\xi^\perp\,,\,N_i=g_{ij} \xi^j)$ and hence 
to a decomposition of the Hamiltonian generator
for Poincar\'e transformations in phase space. 
In particular $H[1,0]$ coincides with the ordinary Hamiltonian $H$ generating proper time evolution
away from the initial hyperplane.
For the explicit value and the derivation of 
$(\xi^\perp,\vect{\xi})$ see Section 3 of~\cite{Tanzi:2020} or Section 2 of~\cite{Henneaux-ED}.
For the following discussion, it suffices to know that, in radial-angular coordinates, they are explicitly
\begin{equation}
    \xi^{\perp}=r b (\barr{x}) +T \,, \qquad
    \xi^r = W (\barr{x}) \,, \qquad
    \xi^{\bar a} = Y^{\bar a} (\barr{x}) + \frac{1}{r} \barr{\gamma}^{\bar a \bar m}\, \partial_{\bar m} W (\barr{x}) \,,
\end{equation}
where $b (\barr{x})$ is responsible for the Lorentz boosts and is an odd function of the sphere, $T$ is responsible for the time translations and is a constant, $W (\barr{x})$ is responsible for the spatial translations, and $Y (\barr{x})$ is the Killing vector field of the metric of the unit two-sphere $\barr{\gamma}$ and is responsible for the rotations.

Before we actually derive the $H[N,\vect{N}]$, let us note three facts.
First, the transformation of the fields under tangential hypersurface deformations (parametrised by $\vect{N}$) can be determined from geometrical considerations.
Specifically, one needs merely to require that the tangential transformations are given by Lie derivatives.
Therefore, the ensuing discussion and computations would be symplified by setting $\vect{N} = 0$.
Nevertheless, we prefer to present the full derivation leaving an arbitrary shift $\vect{N}$ in the remainder of this section.
For comparison, see~\cite[Sec. 3]{Kuchar-Stone}, where $H[N, \vect{N}]$ is derived in the case of free electrodynamics on a spacetime manifold $M = \real \times \Sigma$, being $\Sigma$ a three-dimensional closed manifold.
Those results can be readily applied to our situations up to boundary terms, which are trivially absent in~\cite{Kuchar-Stone}.
It is worth noting already at this point that obstructions to a well-defined Hamiltonian action of the Poincar\'e group are usually caused by the boost in the orthogonal deformation~$\xi^\perp$, so that one should usually pay more attention to the contribution due to $N$ rather than the one due to $\vect{N}$.

Secondly, although we are dealing with flat Minkowski spacetime, it is more convenient to leave the three-metric $g$ in general coordinates for now.
Later on, we will express it in radial-angular coordinates, but there is no advantage in doing it at this stage.
From now on, spatial indices are lowered and raised using the three-metric $g$ and its inverse.

Thirdly, the complex scalar field $\varphi$ can be decomposed into
\begin{equation}
 \varphi = \frac{1}{\sqrt{2}} (\varphi_1 + i \varphi_2)
\end{equation}
where $\varphi_1$ and $\varphi_2$ are two real scalar fields.
Although this replacement makes some expansion less compact, it also makes clearer which are the actual degrees of freedom, with respect to which we have to vary the action.
In the following discussion, we will express the results either in terms of the complex scalar field $\varphi$ or in terms of the two real scalar fields $\varphi_1$ and $\varphi_2$, depending on which of the two approaches is more convenient in each situation.

The action~(\ref{action-Lagrangian}) becomes $S=\int dt \, L[A,\dot A,\varphi,\dot\varphi;g,N,\vect{N}]$, where the Lagrangian is
% \begin{equation} \label{Lagrangian}
% \begin{aligned}
%   L[A,\dot A,\varphi,\dot\varphi;g,N] &=\int d^3 x N \sqrt{g} \left\{
%   \frac{1}{2 N^2} g^{ab} F_{0a}F_{0b}
%   -\frac{1}{4} F_{ab} F^{ab} + \right. \\
%   & +\frac{1}{2 N^2} g^{ab} \Big[ (\dot{\varphi}_1-A_0 \varphi_2)^2 + (\dot{\varphi}_2+A_0 \varphi_1)^2  \Big] - V(\varphi^* \varphi) +\\
%   & \left. - \frac{1}{2} g^{ab} \Big[ (\partial_a \varphi_1 - A_a \varphi_2)(\partial_b \varphi_1 - A_b \varphi_2) +
%   (\partial_a \varphi_2 + A_a \varphi_1)(\partial_b \varphi_2 + A_b \varphi_1) \Big]
%   \right\} +\\
%   &+(\text{boundary}) \,.
% \end{aligned}
% \end{equation}
\begin{equation} \label{Lagrangian}
\begin{aligned}
  L%[A,\dot A,\varphi,\dot\varphi;g,N,\vect{N}]
  ={}& \int d^3 x N \sqrt{g} \left\{
  \frac{1}{2 N^2} g^{ab} F_{0a}F_{0b}
  + \frac{g^{ab} N^c}{ N^2 } F_{0a} F_{bc}
  -\frac{1}{4} F_{ab} F^{ab}
  + \frac{ g^{ac} N^b N^d}{2 N^2} F_{ab} F_{cd} + \right. \\
  & +\frac{1}{2 N^2} \Big[ (\dot{\varphi}_1-A_0 \varphi_2)^2 + (\dot{\varphi}_2+A_0 \varphi_1)^2  \Big]  +\\
  & - \frac{N^a}{N^2} \left[
  (\dot \varphi_1  - A_0 \varphi_2)(\partial_a \varphi_1 -A_a \varphi_2)
  + (\dot \varphi_2 +A_0 \varphi_1)( \partial_a \varphi_2 + A_a \varphi_1)
  \right] +  \\
  & - \frac{1}{2} \left( g^{ab} -\frac{N^a N^b}{N^2} \right)
  \Big[ (\partial_a \varphi_1 - A_a \varphi_2)(\partial_b \varphi_1 - A_b \varphi_2) +
  (\partial_a \varphi_2 + A_a \varphi_1)(\partial_b \varphi_2 + A_b \varphi_1) \Big]
   +\\
  &  - V(\varphi^* \varphi) \bigg\}+(\text{boundary terms}) \,.
\end{aligned}
\end{equation}
First of all, let us note that the above expression does 
not contain any time derivative of $A_0$. This leads to 
a primary constraint expressing that the conjugate 
momentum to $A_0$ has to vanish. A secondary constraint 
follows from the requirement that the conjugate momentum 
to $A_0$ has to stay zero under Hamiltonian evolution. 
According to the standard scheme \cite{Dirac-book}, 
this secondary constraint is then added with a Lagrange 
multiplier to the Hamiltonian, which in our case has 
the simple effect to just add this Lagrange multiplier 
to $A_0$. Now, since the primary constraint tells us 
that the canonical pair consisting of $A_0$ and its 
conjugate momentum do not correspond to physical degrees 
of freedom and therefore should be removed from phase 
space, we can, in our case, shortcut this two-step 
procedure by just saying that $A_0$ ceases to be a 
canonical variable and turns into a mere Lagrange multiplier 
(thereby implementing the secondary constraint). This will 
be seen explicitly in the next section. There 
are no tertiary or higher constraints associated
to $A_0$ in our case. Systematic expositions of
this and similar cases may, e.g.,  be found in
\cite{Kuchar-Stone} and \cite{Kuchar:1976c}.

Secondly, the variation of the Lagrangian~(\ref{Lagrangian}) 
with respect to $\dot A_a$ yields the conjugate three-momenta
\begin{equation} 
\label{momenta}
 \pi^a \eqdef \frac{\delta L}{\delta \dot A_a} 
=\frac{\sqrt{g}}{N} \, g^{ab} 
\big( F_{0b} + N^m F_{bm} \big) \,,
\end{equation}
which are vector densities of weight $+1$.
% and the primary constraints
% \begin{equation} \label{primary-constraints}
%  \pi^0 \eqdef \frac{\delta L}{\delta \dot A_0} \weq 0 \,.
% \end{equation}
Thirdly, the variation of the Lagrangian with respect 
to the real scalar fields gives the further three-momenta
\begin{align} 
 \label{momenta-scalar1}
 \Pi_1 \eqdef& \frac{\delta L}{\delta \dot{\varphi}_1} =
 \frac{\sqrt{g}}{N} \Big[\dot{\varphi}_1 - A_0 \varphi_2
 -N^m (\partial_m \varphi_1 - A_m \varphi_2)\Big] 
 \qquad \text{and}
 \\
 \label{momenta-scalar2}
 \Pi_2 \eqdef& \frac{\delta L}{\delta \dot{\varphi}_2} =
 \frac{\sqrt{g}}{N} \Big[\dot{\varphi}_2 + A_0 \varphi_1
 -N^m (\partial_m \varphi_2 + A_m \varphi_1)\Big] \,,
\end{align}
which are scalar densities of weight $+1$ and can be rewritten in the more compact form
\begin{equation}  \label{complex-field-momentum}
    \Pi \eqdef \frac{1}{\sqrt{2}} \big( \Pi_1 + i \Pi_2 \big) =
    \frac{\sqrt{g}}{N} \Big[D_0 \varphi -N^m D_m \varphi \Big] \,.
\end{equation}

Finally, the symplectic form, from which the Poisson brackets ensue, is
\begin{equation} \label{symplectic-form}
 \Omega[A_a,\pi^a,\varphi,\Pi]=
 \int d^3 x \Big( \extder \pi^a \wedge \extder A_a
 + \extder \Pi_1 \wedge \extder \varphi_1
 + \extder \Pi_2 \wedge \extder \varphi_2
 \Big)
 +(\text{boundary terms}),
\end{equation}
where the bold $\extder$ and $\wedge$ are, respectively, the exterior derivative and the wedge product in phase space.
Note that we are allowing the standard symplectic form to be complemented by a boundary term, which could emerge as a consequence of the boundary term included in the action.
A detailed discussion about boundary terms will be done in the next sections.
Before we complete the derivation of the generator $H[N, \vect{N}]$ and provide its explicit expression, let us briefly discuss the consequences of treating $A_0$ as a Lagrange multiplier.

\subsection{ Constraints and constraints' algebra}
As explained above, in our case, the combined
consequence of the fact that no time derivative
of $A_0$ appears in the action (primary
constraint) and that the removal of $A_0$ and 
its conjugate momentum from phase space 
be consistent with Hamiltonian time evolution 
(secondary constraint) is that we may 
just disregard $A_0$ as a canonical variable 
and turn it into a Lagrange multiplier, 
the variation of which implies the only 
remaining (secondary) constraint. Following 
the standard terminology \cite{Dirac-book}, 
this constraint is of the form 
\begin{equation} 
\label{gauss-constraint}
\mathscr{G} \eqdef \partial_a \pi^a 
+ \varphi_1 \Pi_2 - \varphi_2 \Pi_1 \weq 0 \,,
\end{equation}
where the ``weak equality'' sign $\weq$ 
is meant to express that physical states 
are restricted to those points in phase 
space obeying that equality and also that 
points connected by the Hamiltonian flow generated by $\mathscr{G}$ are considered physically indistinguishable. In particular, 
the dynamical evolution must take place in 
that subset. 

In theory, in order for the Hamiltonian picture
to be consistent, the constraints need to be
preserved by time evolution --- up to 
constraints --- and, in the case of Poincar\'e
invariant theories, by Poincar\'e 
transformations as well!
If this does not happen, one needs to proceed
following the Dirac algorithm~\cite{Dirac-book}.

In practice, this is readily satisfied in 
our case. Indeed, although we have not given 
the explicit value of the Hamiltonian yet, 
let us anticipate that the variation of 
the Gauss constraint is
\begin{equation}
 \delta \mathscr{G}  = \{ \mathscr{G}, H [N, \vect{N}] \} = 0 \,.
\end{equation}
This shows that we have found all the 
constraints of the theory, namely the 
Gauss constraint $\mathscr{G}$, as this 
is preserved not only by time evolution, 
but also by the Poincar\'e transformations.

Finally, one can trivially verify that the constraint is first class and, more precisely, satisfy the abelian algebra
\begin{equation}
\label{constraint-algebra}
 %\{ \pi^0 (x), \pi^0 (x') \}=0 \,, \qquad
 %\{ \pi^0 (x), \mathscr{G} (x') \}=0 \,,\qquad 
 \{ \mathscr{G} (x), \mathscr{G} (x') \}=0 \,.
\end{equation}

Having determined all the constraints of the theory, we can now complete the derivation of  $H[N, \vect{N}]$, which includes both the Hamiltonian and the generator of the Poincar\'e transformations.

\subsection{Hamiltonian, equations of motion, and Poincar\'e transformations}
The generator $H[N,\vect{N}]$ can be readily obtained by means of the Legendre transformation of the Lagrangian, i.e. 
$H \eqdef \int d^3 x \, (\pi^\alpha \dot{A}_\alpha + \Pi_1 \dot{\varphi}_1 +\Pi_2 \dot{\varphi}_2)-L$, in which one replaces $\dot A_a$ with $\pi^a$ by means of~(\ref{momenta}) and $\dot{\varphi}_{1,2}$ with $\Pi_{1,2}$ by means of~(\ref{momenta-scalar1}) and~(\ref{momenta-scalar2}), respectively.
In addition, following~\cite[Sec. 3]{Kuchar-Stone} up to a sign, we define
\begin{equation} \label{A-perp}
    A_\perp \eqdef \frac{1}{N} (A_0 - N^m A_m )
\end{equation}
At this point, we are finally able to write down explicitly the generator of hypersurface  deformations
\begin{equation} 
\label{Hamiltonian}
H[A,\pi,\varphi,\Pi;g,N,\vect{N};A_\perp]=
 \int  d^3 x \Big[ 
 N \mathscr{H}
 + N^i \mathscr{H}_i
 \Big] + (\text{boundary terms})\,,
\end{equation}
where
\begin{equation} 
\begin{aligned} \label{superHamiltonian-new}
 \mathscr{H} \eqdef& 
 \frac{\pi^a \pi_a + \Pi_1^2 + \Pi_2^2}{2\sqrt{g}}
 + \frac{\sqrt{g}}{4} F_{ab} F^{ab}
 -A_\perp \, \mathscr{G}
 + \frac{\sqrt{g}}{2} g^{ab} \big( \partial_a \varphi_1 \partial_b \varphi_1 + \partial_a \varphi_2 \partial_b \varphi_2 \big) + \\
 & + \sqrt{g} A^a \big( \varphi_1 \partial_a \varphi_2 - \varphi_2 \partial_a \varphi_1 \big)
 +\frac{1}{2} A_a A^a \big( \varphi_1^2 + \varphi_2^2 \big)
 + \sqrt{g} \, V(\varphi^* \varphi)
\end{aligned}
\end{equation}
is responsible for the orthogonal transformations and
\begin{equation} 
 \mathscr{H}_i \eqdef
 \pi^a\partial_i A_a - \partial_a (\pi^a A_i)
 + \Pi_1 \partial_i \varphi_1 + \Pi_2 \partial_i \varphi_2
\end{equation}
is responsible for the tangential transformations.
Note that the Gauss constraint~(\ref{gauss-constraint}) appears in the generator~(\ref{Hamiltonian}) multiplied by the Lagrange multiplier $A_\perp$.

The knowledge of the symplectic form~(\ref{symplectic-form}) and of the generator of hypersurface deformations~(\ref{Hamiltonian}) allows us to determine how the fields vary infinitesimally under said deformation.
Specifically, the infinitesimal change of the fields under a transformation is represented by a vector field $X = (\delta A_a \,, \delta \pi^a \,, \dots)$ in phase space.
Knowing the symplectic form~(\ref{symplectic-form}) and the generator~(\ref{Hamiltonian}), the corresponding vector field can be determined by means of the equation $\extder H = - i_X \Omega$, which, in general, might be impossible to fulfil due to the presence of boundary terms in the variation $\extder H$ which cannot be compensated by the inclusion of boundary terms in the symplectic structure.
If we neglect this issue for the moment, as it will be thoroughly discussed in the next sections, we find
\begin{align} \label{eoms-begin}
 \delta A_a ={}& N \frac{\pi_a}{\sqrt{g}} 
 + \partial_a (N A_\perp ) + \mathscr{L}_{\vect{N}} A_a \,,\\
 \delta \pi^a ={}& \partial_b (N \sqrt{g}\, F^{ba})
    - 2 \sqrt{g} \, N \,\imaginarypart \left( \varphi^* D^a \varphi \right) 
    + \mathscr{L}_{\vect{N}} \pi^a\,, \\
 \delta \varphi ={}& N \frac{\Pi}{\sqrt{g}}
 - i (N A_\perp) \varphi +\mathscr{L}_{\vect{N}} \varphi \,, \\
 \label{eoms-end}
 \delta \Pi ={}& D^a ( \sqrt{g} \, N D_a \varphi )
 + \sqrt{g} \, N \Big( \mu^2 - 2 \lambda |\varphi|^2  \Big) \varphi
 - i ( N A_\perp ) \Pi +\mathscr{L}_{\vect{N}} \Pi \,,
\end{align}
where $\mathscr{L}_{\vect{N}}$ is the three-dimensional Lie derivative with respect to $N^i$ and we have chosen to use the more compact complex notation.
The above equations reduce to the equations of motion when $N = 1$ --- in which case the left-hand sides become the time derivative of the fields --- and to the Poincar\'e transformations when $N = \xi^\perp = r b (\barr{x}) + T$ and $N^i = \xi^i$.

\subsection{Gauge transformations} \label{subsec:gauge-transformations}
The presence of the Gauss constraints~(\ref{gauss-constraint}) in the Hamiltonian~(\ref{Hamiltonian}) causes the transformations~(\ref{eoms-begin})--(\ref{eoms-end}) to include a gauge transformation, whose gauge parameter is the arbitrary function $\zeta \eqdef - N A_\perp$.
In order to ensure the uniqueness of solutions despite the arbitrariness of $\zeta$, one needs to treat this transformations as mere relabelling of a physical state, i.e., a redundancy in the mathematical description of the theory.

The infinitesimal form of the gauge transformations, which we can read from
 the transformations~(\ref{eoms-begin})--(\ref{eoms-end}), is
\begin{equation} \label{gauge-infinitesimal}
 \delta_\zeta A_a = - \partial_a \zeta \,,\qquad
 \delta_\zeta \pi^a = 0 \,, \qquad
 \delta_\zeta \varphi = i \zeta \varphi \,, \qquad \text{and} \qquad
 \delta_\zeta \Pi = i \zeta \Pi \,.
\end{equation}
Specifically, these are generated by
\begin{equation} \label{gauge-generator}
    G[\zeta] = \int d^3 x \, \zeta(x) \mathscr{G} (x)
\end{equation}
through the equation $\extder G[\zeta] = - i_{X_\zeta} \Omega$.
The left-hand side of this equation can be readily computed to be
\begin{equation}
\begin{aligned}
 \extder G[\zeta] ={}& \int d^3 x \, \Big[
   -\partial_a \zeta \extder \pi^a
   +\zeta \Pi_2 \extder \varphi_1
   -\zeta \Pi_1 \extder \varphi_2
   -\zeta \varphi_2 \extder \Pi_1
   +\zeta \varphi_1 \extder \Pi_2
   \Big] + \\ 
   & + \lim_{R \rightarrow \infty} \oint_{S^2_R} d^2 \barr{x}_k  \, \zeta \extder \pi^k \,.
\end{aligned}
\end{equation}
Assuming that the symplectic form~(\ref{symplectic-form}) does not contain any boundary term, the equation $\extder G[\zeta] = - i_{X_\zeta} \Omega$ is fulfilled so long as the boundary term in the above expression is actually zero.
Whether or not this is the case, and for which class of functions $\zeta (x)$ this happens, vastly depends on the asymptotic behaviour of the fields.
We will discuss this in the next sections and we will see that the asymptotic behaviour of the fields changes depending on the choice of parameters in the potentials~(\ref{potential}), i e., on whether we are dealing with scalar electrodynamics or with the abelian Higgs model.
For now, let us note that the generator $G[\zeta]$ can be in general extended to
\begin{equation} \label{gauge-generator-improper}
    G_{\text{ext.}}[\zeta] = G[\zeta] - \lim_{r \rightarrow \infty} \oint_{S^2_r} d^2 \barr{x}_k  \, \zeta \pi^k 
\end{equation}
whose variation, now, does not contain any boundary term.
In the case in which the boundary term in the above expression is non-trivial, the transformations corresponding to $G_{\text{ext.}}[\zeta]$ are not, in fact, proper gauge transformations.
Rather they are true symmetries of the theory relating physically-different states and are commonly referred to as \emph{improper gauge transformations}, following~\cite{Teitelboim-YM2}. 
In addition, one can define the charge
\begin{equation} \label{charge}
    Q[\zeta] \eqdef
    \lim_{r \rightarrow \infty} \oint_{S^2_r} d^2 \barr{x}_k  \, \zeta \pi^k \,,
\end{equation}
which implies $G_{\text{ext.}}[\zeta] = G[\zeta] - Q[\zeta] \approx -Q[\zeta]$, so that one can tell whether a transformation is a proper gauge or an improper one by checking whether the charge is zero or not, respectively.\footnote{
Note that, due to the limit in the definition~(\ref{charge}), the charge depends only on the asymptotic values of the fields and of the gauge parameter $\zeta$, which are going to be thoroughly discussed in the next sections.
Let us anticipate that the asymptotic part of $\zeta(x)$ is going to be denoted by $\barr{\zeta}(\barr{x})$, which is a function on the two-sphere at infinity.
Thus, the charge can be written simply as $Q \left[ \,\barr{\zeta}\, \right]$, which is usually decomposed into spherical-harmonics components.
Specifically, one defines $Q_{\ell m} \eqdef Q[Y_{\ell m}]$ in terms of the spherical harmonics $Y_{\ell m}$.
The component $Q_{00}$ corresponds to the global (electric) charge.
}
Whether or not there is a non-trivial class of functions $\zeta(x)$ such that improper gauge transformations exist and have a well-defined action on phase space depends, again, on the asymptotic behaviour of the fields, which will be discussed in the next sections.

Finally, let us point out that the expressions for the infinitesimal gauge transformations~(\ref{gauge-infinitesimal}) can be integrated to get the finite form of gauge transformations
\begin{equation} \label{gauge-finite}
 \Gamma_\zeta ( A_a ) = A_a - \partial_a \zeta \,,\qquad
 \Gamma_\zeta ( \pi^a ) = \pi^a \,, \qquad
 \Gamma_\zeta ( \varphi ) = e^{i \zeta} \, \varphi \,, \qquad \text{and} \qquad
 \Gamma_\zeta ( \Pi ) = e^{i \zeta} \, \Pi \,,
\end{equation}
where $\Gamma_\zeta$ denotes the action of $e^{i \zeta (x)} \in \U(1) $ on the fields, i.e. $A_a \mapsto A_a' = \Gamma_\zeta (A_a)$ for instance.
This clearly show the $\U(1)$ nature of the gauge symmetry.

In the next sections, we are going to discuss the specific cases of scalar electrodynamics and of the abelian Higgs model.
Before that, in the next section, we are going to briefly discuss the case of a free scalar field with the potential~(\ref{potential}), as this simple situation let us highlight some of the features of the asymptotic structure.

\section{Free scalar field} 
\label{sec:free-scalar}
Let us study first the behaviour of the complex scalar field when it is not coupled to the gauge potential.
This can be achieved by considering the equations~(\ref{eoms-begin})--(\ref{eoms-end}) and setting to zero the values of the gauge potential $A_a$, the conjugated momenta $\pi^a$, and the Lagrange multiplier $A_\perp$, obtaining
\begin{align} \label{eoms-complex-begin}
 \delta \varphi ={}& N \frac{\Pi}{\sqrt{g}} +\mathscr{L}_{\vect{N}} \varphi \,, \\
 \label{eoms-complex-end}
 \delta \Pi ={}&  \nabla^a ( \sqrt{g} N \partial_a \varphi ) + N \sqrt{g} \Big( \mu^2 - 2\lambda \, |\varphi|^2 \Big) \varphi
 + \mathscr{L}_{\vect{N}} \Pi \,.
\end{align}
Finally, the generator of hypersurface deformations~(\ref{Hamiltonian}) reduces to
\begin{equation} \label{Hamiltonian-scalar}
 \begin{aligned}
  H_{\text{scalar}}[\varphi,\Pi;g,N,\vect{N}]={}&
  \int d^3 x \, \left\{
  N \left[
  \frac{\Pi^2}{\sqrt{g}}
  +\sqrt{g} g^{ab} \partial_a \varphi^* \partial_b \varphi
  + \sqrt{g} \Big( -\mu^2 |\varphi|^2 + \lambda |\varphi|^4 \Big)
  \right] + \right. \\
  &+  2 N^i \,\realpart \big( \Pi^* \partial_i \varphi \big)  \Big\} 
  +(\text{boundary terms}) \,,
 \end{aligned}
\end{equation}

Let us analyse separately the two different scenarios in the next two subsections.
First, we will consider the case in which $m^2 \eqdef - \mu^2 \ge 0$ and $\lambda \ge 0$.
This describes a massive ($m^2 >0$) or massless ($m^2 = 0$) complex scalar field with ($\lambda > 0$) or without ($\lambda = 0$) a self interaction.
This will be useful when studying scalar electrodynamics in section~\ref{sec:scalar-electrodynamics}.
Secondly, we will consider the case in which $\mu^2 > 0$ and $\lambda > 0$, so that the potential takes the well-known Mexican-hat shape.
This will be relevant in the analysis of the abelian Higgs model in section~\ref{sec:abelian-Higgs}.

\subsection{Massless and massive scalar field} \label{subsec:massless-massive-scalar}
Let us first consider the case of a scalar field with squared mass $m^2 \eqdef - \mu^2 \ge 0$.
The self interaction is either present or not, i.e., $\lambda \ge 0$.
Note that the equation of motion~(\ref{eoms-complex-begin})--(\ref{eoms-complex-end}) contain the trivial solution
\begin{equation}
 \varphi^{(0)} (x) = 0
 \qquad \text{and} \qquad
 \Pi^{(0)} (x) =0 \,,
\end{equation}
which is also the solution that minimises the potential~(\ref{potential}) and the energy.
Indeed, neglecting the boundary, the value of the Hamiltonian for this solution is
\begin{equation}
 E^{(0)} \eqdef H_{\text{scalar}}[\varphi^{(0)},\Pi^{(0)};g,N=1, \vect{N} = 0] = 0 \,,
\end{equation}
whereas the value of the Hamiltonian for any other field configuration is positive.

At this point, we use a power-like ansatz for the fall-off behaviour of the field and the potential.
In detail, we assume that they behave as
\begin{equation}
 \varphi (x) = \frac{1}{r^\alpha} \barr{\varphi} (\barr{x}) +\bigo \big( 1/r^{\alpha+1} \big)
 \qquad \text{and} \qquad 
 \Pi (x) = \frac{1}{r^\beta} \barr{\Pi} (\barr{x}) +\bigo\big( 1/r^{\beta+1} \big)
\end{equation}
in radial-angular coordinates.
Whether or not $\alpha$ and $\beta$ can be found, such that the fall-off conditions are preserved by the Poincar\'e transformations, depends crucially on the value of the mass.
More precisely, in the \emph{massless} case, i.e. $m=0$, one finds the fall-off conditions
\begin{equation} \label{fall-off-scalar-massless}
 \varphi_\text{massless} (x) = \frac{1}{r} \barr{\varphi} (\barr{x}) +\bigo\big( 1/r^2 \big)
 \qquad \text{and} \qquad 
 \Pi_\text{massless} (x) = \barr{\Pi} (\barr{x}) +\bigo\big( 1/r \big) \,,
\end{equation}
which also make the symplectic form logarithmically divergent.
%Note that, although the fall-off conditions above were derived from the requirement that they are preserved by the orthogonal part of the Poincar\'e transformations, one can easily verify that they are preserved also by their tangential part.

Before we discuss the massive case, let us point out that, in order to make the symplectic form actually finite, one needs to impose parity conditions on the asymptotic part of the fields in addition to the aforementioned fall-off conditions.
Specifically, it suffices, for instance, to require that $\barr{\varphi} (\barr{x})$ is either an even or an odd function of the sphere under the antipodal map\footnote{
We remind that the antipodal map, denoted hereafter by $\Phi \colon\barr{x} \mapsto - \barr{x}$, consists in the explicit transformation
$(\theta, \phi) \mapsto (\pi - \theta, \phi+\pi)$ in terms of the standard spherical coordinates.
A generic tensor field $T$ (or a density) is said to be \emph{even} under the antipodal map if $\Phi^* T = T$, being $\Phi^*$ the pull back of the antipodal map.
Analogously, $T$ is \emph{odd} if $\Phi^* T = -T$.
To see how this translate into the exact parity of the components of a tensor field (or density) expressed in some coordinates like the standard $(\theta,\phi)$ spherical coordinates, see footnote 2 of~\cite{Henneaux-ED}.
}
and, at the same time, that $\barr{\Pi}(\barr{x})$ has the opposite parity.
In this way, the potentially logarithmically-divergent term in the symplectic form is, in fact, zero.
It is easy to check that these parity conditions are preserved by the Poincar\'e transformations, which take the asymptotic form
\begin{align}
 \delta_{\xi} \barr{\varphi} &= \frac{b \, \barr{\Pi}}{\sqrt{\barr{\gamma}}} + Y^{\bar m} \partial_{\bar m} \barr{\varphi} \,,
 \\
 \delta_{\xi} \barr{\Pi} &= - b \sqrt{\barr{\gamma}} \, \barr{\varphi}
 + \barr{\nabla}^{\bar m} \left( \sqrt{\barr{\gamma}} \, b \partial_{\bar m} \barr{\varphi} \right)
 -2 b \sqrt{\barr{\gamma}} \, \lambda |\barr{\varphi}|^2 \, \barr{\varphi}
 + \partial_{\bar m} \left( Y^{\bar m} \barr{\Pi} \right) \,,
\end{align}
where $\barr{\nabla}$ denotes the covariant derivative of the round unit sphere, $b$ parametrises the Lorentz boost, and Killing vector field $Y$ of the round-unit-two-sphere metric $\barr{\gamma}$ parametrises the rotations.
We will come back to the discussion about parity conditions in section~\ref{sec:scalar-electrodynamics} where we will consider the couple of the scalar field to electrodynamics.

In the massive case, the appearance of a new term proportional to $m^2 > 0$ in the Poincar\'e transformations of the momentum, substantially modifies the fall-off behaviour of the fields, so that one does not find any power-like  solution.
One can show, as it is done in appendix~\ref{appendix:massive-fall-off}, that both $\varphi$ and $\Pi$ need to be function approaching zero at infinity faster than any power-like function.\footnote{
We will refer to this fall-off behaviour of the scalar field and its momentum and to similar behaviours encountered in the remainder of this paper by saying that the fields are ``quickly vanishing (at infinity)''.
}
Hence, we will restrict the phase space by requiring that both $\varphi$ and $\Pi$ are quickly-falling functions.
% Schwartz functions.\footnote{
% We recall a smooth function belongs to the Schwartz space if, and only if, the function itself and any of its (partial) derivatives falls off at infinity faster than any power-like function.
% In symbols,
% \[
% \Schwartz(\real^n) \eqdef
% \left\{
% f \in C^{\infty}(\real^n) \colon
% \forall (\alpha_1,\dots,\alpha_n) \in \naturals^n \,, \forall \beta \in \naturals \,, \;
% \lim_{r \rightarrow \infty} \left( r^\beta \, \partial_1^{\alpha_1} \cdots \partial_n^{\alpha_n} f \right) = 0
% \right\} \,,
% \]
% where $\partial_k^{\alpha_k}$ denotes the partial derivative of $\alpha_k$-th order with respect to $x^k$ and $r \eqdef |x|$.
% In addition, we say that a tensor fields (or density) on a flat spacetime belongs to $\Schwartz$ if each one of its components in Cartesian coordinates is a Schwartz function. 
% }
% In theory, we could consider a function space slightly bigger than $\Schwartz$, since we do not need to control on every derivative order, as we do in $\Schwartz$, but only up to second order for $\varphi$ according to the Poincar\'e transformations.
% In practice, this does not affect significantly the potentially-interesting solutions to the equations of motion, while increasing the regularity of these solutions and gaining some useful mathematical properties.
In details, we will require that the scalar field $\varphi$, as well as its spatial derivatives up to second order, and the momentum $\Pi$ vanish in the limit to spatial infinity faster than any power-like functions (in Cartesian coordinates).
Note that, due to these fall-off conditions, the Hamiltonian and the generator of the Poincar\'e transformations of the massive scalar field are finite and functionally differentiable with respect to the canonical fields without any need of a boundary term.

Finally, let us note that the theory, both in the massless and in the massive case, possesses the global $\U (1)$ symmetry
\begin{equation} \label{global-symmetry-transformations}
 \big[ \Gamma_\zeta ( \varphi ) \big] (x) = e^{i \zeta} \, \varphi (x) 
 \qquad \text{and} \qquad
 \big[ \Gamma_\zeta ( \Pi ) \big] (x) = e^{i \zeta} \, \Pi (x) \,,
\end{equation}
where $\Gamma_\zeta$ denotes the action of $e^{i \zeta} \in \U (1)$ on the fields.
Note that, differently from~(\ref{gauge-finite}), the action is that of the global $\U (1)$, i.e., the parameter $e^{i \zeta} \in \U (1)$ is the same at each spacetime point.
The infinitesimal version of the above transformations is generated by
\begin{equation} \label{global-symmetry-generator}
    G[\zeta] = \int d^3 x \, \zeta \Big[
    \varphi_1 (x) \Pi_2 (x) - \varphi_2 (x) \Pi_1 (x)
    \Big] \,,
\end{equation}
where, again, $\zeta$ is independent of $x$.
Note that the above generator is always finite and differentiable, i.e. $\extder G[\zeta] = -i_{X_\zeta} \Omega$.\footnote{
In the massless case, the generator is finite thanks to the combination of the fall-off and parity conditions.
The former alone would make the generator logarithmically divergent.
}
In additions, it is \emph{not} proportional to a constraint, as the theory of a free scalar field (with a potential) does not possess any constraint.
It can be easily verified that the generator above Poisson-commutes with Hamiltonian, i.e.
\begin{equation}
\Big\{ G[\zeta] , H[N=1,\vect{N} = 0] \Big\} \eqdef
i_{X_G} \big( i_{X_H} \Omega \big) =  0 \,,    
\end{equation}
showing that it generates indeed a symmetry.

To sum up, in this subsection, we have studied the fall-off conditions of a complex scalar field and its conjugated momentum with or without a quartic self-interaction.
The fall-off behaviour of the field and the momentum crucially depends on whether or not the mass is zero.
On the one hand, in the massless case, the fall-off conditions are power-like and, precisely, the ones in~(\ref{fall-off-scalar-massless}).
% One can show that a massless scalar field coupled to electrodynamics leads to some non-trivial asymptotic symmetries and to some issues as well.
%We will discuss the details of this in section~\ref{sec:scalar-electrodynamics},
On the other hand, in the massive case, the scalar field --- as well as its spatial derivatives up to second order --- and its momentum need to vanish at spatial infinity faster than any power-like function.
% , in particular, are assumed to belong to the Schwartz space, with an only-very-slight loss of generality.
% As a consequence, any charges associated to any possible asymptotic symmetry vanishes and the only symmetry left is the Poincar\'e group.
% , as we do in~\ref{appendix:massive-null-infinity}.
In addition, we have seen that the theory possesses a global $\U (1)$ symmetry.

\subsection{Mexican-hat potential} \label{subsec:Mexican-hat-potential}
Let us now consider the case of a scalar field with a Mexican-hat potential $\mu^2 \ge 0$ and $\lambda > 0$.
Note that the parameter of the self interaction $\lambda$ has to be strictly positive for, otherwise, the potential and, as a consequence, the Hamiltonian are not bounded from below.
As in the cases of a massive and massless scalar field, the equation of motion~(\ref{eoms-complex-begin})--(\ref{eoms-complex-end}) contain the trivial solution
\begin{equation} \label{solution-zero}
 \varphi^{(0)} (x) = 0
 \qquad \text{and} \qquad
 \Pi^{(0)} (x) =0 \,.
\end{equation}
However, this is not any more the solution that minimises the potential $V(\varphi^* \varphi)$ and the Hamiltonian.
Indeed, this solution is found at a local maximum of the potential and gives the value of the Hamiltonian
\begin{equation}
 E^{(0)} \eqdef H_{\text{scalar}}[\varphi^{(0)},\Pi^{(0)};g,1,0] = 0 \,.
\end{equation}

In this case, the potential and the Hamiltonian are minimised by the constant solutions to the equations of motion
\begin{equation} \label{solutions-ssb}
 \varphi^{(\vartheta)} (x) = \frac{v}{\sqrt{2}} \, e^{i \vartheta}
 \qquad \text{and} \qquad 
 \Pi^{(\vartheta)} (x) = 0 \,,
\end{equation}
where the parameter $\vartheta$ belongs to $\real/2\pi$ and $v \eqdef \sqrt{\mu^2/ \lambda}$.
On all these solutions, neglecting the boundary, the Hamiltonian takes the same value
\begin{equation}
 E^{(\vartheta)} \eqdef H_{\text{scalar}}[\varphi^{(\vartheta)},\Pi^{(\vartheta)};g,1,0] =
 - \int d^3 x \, \sqrt{g} \, \lambda \, \frac{v^4}{4} \,,
\end{equation}
which diverges to $-\infty$, since it is the integral of a negative constant over a spatial slice $\Sigma \sim \real^3$.
This means that we would not be able to include the solutions~(\ref{solutions-ssb}) if we wished to have a well-defined, i.e. finite and functionally-differentiable, Hamiltonian.

The solution to this issue is quite simple.
We merely need to redefine the generator of hypersurface deformations~(\ref{Hamiltonian-scalar}) and, as a consequence, the Hamiltonian to
\begin{equation} \label{Hamiltonian-scalar-redefined}
 \begin{aligned}
  H'_{\text{scalar}}[\varphi,\Pi;g,N,\vect{N}] ={}&
 \int d^3 x \, N \left\{
 \frac{\Pi^2}{\sqrt{g}}
 +\sqrt{g} g^{ab} \partial_a \varphi^* \partial_b \varphi
 + \sqrt{g} \Big( \lambda  \frac{v^4}{4}
 -\mu^2 |\varphi|^2 + \lambda |\varphi|^4 \Big)
  +  \right. \\
  &+  2 N^i \,\realpart \big( \Pi^* \partial_i \varphi \big)  \Big\} 
  +(\text{boundary terms}) \,.
 \end{aligned}
\end{equation}
This amounts to nothing else than the addition of the constant $\lambda v^4 / 4$ to the potential, without any impact on the equations of motion and on the Poincar\'e transformations.
Thus, neglecting the boundary, the value of the Hamiltonian evaluated on the solutions~(\ref{solutions-ssb}) is
\begin{equation}
 E'{}^{(\vartheta)} \eqdef H'_{\text{scalar}}[\varphi^{(\vartheta)},\Pi^{(\vartheta)}; g,1,0] =
 0 \,,
\end{equation}
whereas the value is positive for any other field configuration.
Note that, however, the value of the Hamiltonian evaluated on the trivial solution~(\ref{solution-zero}) is now divergent.
As a consequence, we need to remove this solution from the allowed field configuration, but this does not have a huge impact on the physical side, as~(\ref{solution-zero}) is on a local maximum of the potential and, thus, unstable under perturbations.

Let us now discuss the fall-off conditions of the field and its conjugated momentum.
Although most of the discussion does not differ much from the case of the massive scalar field discussed in the previous subsection, there are nevertheless a few subtleties that one should take into consideration.
We will work in radial-angular coordinates.

First, let us focus on the terms in the Hamiltonian~(\ref{Hamiltonian-scalar-redefined}) containing the potential $V(\varphi^* \varphi)$ with the newly-added constant $\lambda v^4 / 4$.
If we wish this part to be finite upon integration, we need to require the absolute value of the field $|\varphi(x)|$ to approach the value $v / \sqrt{2}$ as $r \rightarrow \infty$.
In other words, this means that, if we write
\begin{equation}
 \varphi (x) = \frac{1}{\sqrt{2}} \rho (x) \, e^{i\vartheta (x)} \,,
\end{equation}
then $\rho (x) = v + h (x)$, where $h (x)$ vanishes in the limit $r \rightarrow \infty$.
Note that, in principle, we allow the phase $\vartheta(x)$ to be non-constant.
Nevertheless, we require that it has a well-defined limit $\barr{\vartheta} (\barr{x}) \eqdef \lim_{r \rightarrow \infty} \vartheta (x)$ as a possibly non-constant function on the sphere at infinity.

Secondly, let us note that $\varphi (x)$ is not vanishing in a neighbourhood of spatial infinity, so that we can always write, and it is convenient to do so,
\begin{equation}
    \Pi (x) = \Big(u(x) + i w(x) \Big) \varphi(x) \,,
\end{equation}
where $u(x)$ and $v(x)$ are both real.
From the transformation of $\varphi$, one can easily find the transformations of its absolute value and phase as
\begin{equation} \label{transformations-abs-phase}
    \delta \rho = \rho \, \realpart \left( \frac{\delta \varphi}{\varphi} \right)
    \qquad \text{and} \qquad
    \delta \vartheta = \imaginarypart \left( \frac{\delta \varphi}{\varphi} \right) \,.
\end{equation}
Analogously, the transformations of $u$ and $w$ can be obtained from those of $\Pi$ and $\varphi$, as
\begin{equation} \label{transformations-u-w}
    \delta u = \realpart \left( \frac{\varphi \, \delta \Pi - \Pi \, \delta \varphi}{\varphi^2} \right)
    \qquad \text{and} \qquad
    \delta w = \imaginarypart \left( \frac{\varphi \, \delta \Pi - \Pi \, \delta \varphi}{\varphi^2} \right) \,.
\end{equation}

Thirdly, we can show that $h (x)$, $u (x)$, and $w (x)$ need to fall off at infinity faster than any power-like functions.
A precise proof of this statement would require us to proceed as in appendix~\ref{appendix:massive-fall-off}.
Instead, let us here provide a less rigorous argumentation.
Specifically, let us assume the power-like behaviours
\begin{equation} \label{power-like-ansatz}
    h(x) = \frac{1}{r^{\alpha}} \barr{h} (\barr{x}) + o(1/r^\alpha) \,, \qquad
    u (x) = \frac{1}{r^{\beta}} \barr{u} (\barr{x}) + o(1/r^\beta) \,, \qquad
    w (x) = \frac{1}{r^{\gamma}} \barr{w} (\barr{x}) + o(1/r^\gamma) \,.
\end{equation}
Note that $\alpha$ need to be greater than zero, since we requested $h$ to vanish as $r$ tends to infinity.
Now, let us consider only a part of the Poincar\'e transformations~(\ref{eoms-begin})--(\ref{eoms-end})
Namely,
\begin{equation} 
 \delta' \varphi = \xi^\perp \frac{\Pi}{\sqrt{g}}
 \qquad \text{and} \qquad
 \delta' \Pi = \sqrt{g} \, \xi^\perp \Big( \mu^2 - 2 \lambda |\varphi|^2  \Big) \varphi \,,
\end{equation}
from which we can derive the corresponding transformations of $h$, $u$, and $v$ using~(\ref{transformations-abs-phase}) and~(\ref{transformations-u-w}), obtaining
\begin{align}
    \delta' h = \frac{\xi^\perp}{\sqrt{g}} u \,, \qquad
    \delta' u =  -2 v \xi^\perp \sqrt{g} \lambda \left( h - \frac{h^2}{2v} \right) - \frac{\xi^\perp}{\sqrt{g}} \big(u^2 -w^2 \big) \,, \qquad
    \delta' w = - \frac{\xi^\perp}{ \sqrt{g}} 2 u w \,.
\end{align}
At this point, we insert~(\ref{power-like-ansatz}) in the above expressions and expand everything in powers of $r$, including $\sqrt{g} = r^2 \sqrt{\barr{\gamma}}$ and $\xi^\perp = r b +T$.
Requiring that the fall-off conditions~(\ref{power-like-ansatz}) are preserved, i.e., that the terms on the right-hand side of the above expressions do not fall off slower than the respective field, we find that the exponents in the power-like ansatz need to satisfy the non-trivial inequalities
\begin{equation} \label{inequalities-alpha-beta-gamma}
    \beta + 1 \ge \alpha \,, \qquad
    \alpha - 3 \ge \beta \,, \qquad
    2 \gamma + 1 \ge \beta \,, \qquad
    \beta + 2 \ge 0 \,,
\end{equation}
where the first inequality comes from the transformation of $h$, the last from that of $w$, and the remaining two from that of $u$.
One sees immediately that the first two inequalities lead to the contradiction
\begin{equation}
    \alpha \le \beta + 1 \le \alpha -2 \,, 
\end{equation}
which would lead to the conclusion that $h$ and $u$ are quickly vanishing at infinity, if one proceeded like in appendix~\ref{appendix:massive-fall-off}.
Furthermore, the third inequality in~(\ref{inequalities-alpha-beta-gamma}) would lead us to the conclusion that also $w$ is quickly vanishing.

Lastly, let us note that the conditions that we have determined so far show us that $\Pi$ and $h$ need to fall-off at infinity faster than any power-like function.
However, we have still to determine the fall-off behaviour of the phase $\vartheta (x)$.
To do so, it suffices to consider the transformation of $\Pi$ under time evolution, i.e., equation~(\ref{eoms-end}) at $N=1$ and $\vect{N}=0$.
Up to terms that are quickly vanishing at infinity, we find
\begin{equation}
    \delta \Pi = \varphi \left[
    - \sqrt{g} \, \partial_a \vartheta \, g^{ab} \, \partial_b \vartheta
    +i \partial_a \left( \sqrt{g} \, g^{ab} \partial_b \vartheta \right)
    \right]
   + (\text{quickly-vanishing terms}) \,.
\end{equation}
Thus, we have to impose that $\partial_a \vartheta$ is quickly vanishing in order to preserve the fall-off condition of $\Pi$.
This leads us to two fact.
First, the asymptotic part $\barr{\vartheta} (\barr{x})$ needs to be constant on the sphere at infinity.
We will simply denote it with $\barr{\vartheta}$.
Second, if we write $\vartheta (x) = \barr{\vartheta} + \chi (x) /v$, we will find out that $\chi (x) $ is quickly vanishing at infinity, as well as its derivatives up to second order.
We will see in section~\ref{sec:abelian-Higgs} that this situation changes when a gauge potential is present, as in the abelian Higgs model.
Finally, note that, from the Poincar\'e transformation of $\vartheta$
\begin{equation}
    \delta \vartheta = \imaginarypart \left( \frac{\delta \varphi}{ \varphi} \right) =
    \imaginarypart  \left( \frac{\xi^\perp \Pi}{\sqrt{g} \varphi}
    +\frac{\mathscr{L}_{\vect{N}} \varphi }{\varphi}\right)  \,,
\end{equation}
we infer that $\barr{\vartheta}$ is invariant under the Poincar\'e transformations and, in particular, is time independent.
Indeed, the first summand on the left-hand side of the above expression is clearly quickly vanishing in the limit $r \rightarrow \infty$, while the second summand reduces to $\mathscr{L}_{\vect{N}} \chi /v$ which, too, is quickly vanishing.

This concludes the derivation of the fall-off conditions of a complex scalar field with a Mexican-hat potential.
In short, we have shown that, when one considers the Mexican-hat potential as in the case of the Higgs mechanism, the generator of hypersurface deformations and the Hamiltonian needs to be modified to~(\ref{Hamiltonian-scalar-redefined}) by adding a constant to the potential, so that the minimal-energy solutions~(\ref{solutions-ssb}) to the equations of motion have finite energy.
The phase space is then defined by all those fields and momenta, whose difference from one of the minimum-energy solutions vanishes at infinity faster than any power-like function.
As in the case of the scalar massive field, one has to require the quick fall-off of the field up to the second-order spatial derivatives.
Note that the asymptotic part of the phase of the scalar field $\barr{\vartheta}$ needs to be constant on the sphere at infinity and is time-independent.
Moreover, the phase $\vartheta$ can differ from its constant value at infinity by a function $\chi/v$ that is quickly vanishing.
This will not be the case when we reintroduce the gauge potential $A_a$, as we shall see in section~\ref{sec:abelian-Higgs}.

Before we move to the study of scalar electrodynamics in section~\ref{sec:scalar-electrodynamics} and to that of the abelian Higgs model in section~\ref{sec:abelian-Higgs}, let us make the connection with the usual interpretation of $h$ and $\chi$ in high-energy physics.
To this end, let us consider the action in the Lagrangian picture, which can be obtained from~(\ref{action-Lagrangian}) by setting $A_\alpha = 0$ and adding the constant $\lambda v^4 /4$ to the potential.
Rewriting this action in terms of $h$ and $\chi$, we obtain
\begin{equation}
    S[h,\chi] = \int d^4 x \left\{
    -\frac{1}{2} \left( {}^4 g^{\alpha \beta} \partial_{\alpha} h \, \partial_\beta h + 2 \mu^2 h^2 \right)
    - \frac{1}{2} {}^4 g^{\alpha \beta} \partial_{\alpha} \chi \, \partial_\beta \chi
    +(\text{interactions})
    \right\} \,,
\end{equation}
where the interactions include all the terms that are not quadratic in the fields.
From the above expression, we read that $h$ is a scalar field of squared mass $m_h^2 \eqdef 2 \mu^2 $, whereas $\chi$ is a massless scalar field.
The latter is precisely the Goldstone boson of the spontaneously broken global $\U (1)$ symmetry.
Indeed, as in the case analysed in the previous subsection, the theory possesses the symmetry~(\ref{global-symmetry-transformations}) generated by~(\ref{global-symmetry-generator}).
However, in this case, the minimum-energy solutions are not invariant under the action of the symmetry.
Rather, the vacuum solution
$\big( \varphi^{(\vartheta)} \,, \Pi^{(\vartheta)}  \big)$ is mapped to the different, physically-nonequivalent vacuum solution
$\big( \varphi^{(\vartheta+\zeta)} \,, \Pi^{(\vartheta+\zeta)}  \big)$ under the action of $\zeta \in \U (1)$.
In the abelian Higgs model analysed in section~\ref{sec:abelian-Higgs}, the Goldstone boson $\chi$ will turn out to be pure gauge, i.e. physically irrelevant, whereas $h$ will be the Higgs boson.

\section{Scalar electrodynamics} 
\label{sec:scalar-electrodynamics}

In this section, we will discuss the asymptotic symmetries of scalar electrodynamics, that is the case of a complex scalar field minimally coupled to electrodynamics.
Specifically, this amount to consider the Hamiltonian~(\ref{Hamiltonian}) in the case in which the parameters in the potential~(\ref{potential}) are such that $m^2 \eqdef - \mu^2 \ge 0$ and $\lambda \ge 0$.
The former parameter represent the (squared) mass of the scalar field and distinguishes between the massive case ($m^2 > 0$) from the massless one ($m^2 = 0$).
The latter parameter regulates the magnitude of the self-interaction of the scalar field and is allowed, in principle, $\lambda$ to be different from zero.

The ensuing discussion vastly differs depending on whether the scalar field is massive or massless.
Therefore, we will keep separated the analyses of these two different situations.
We will begin our discussion with the massive case, as this is significantly simpler and we will dedicate to it the first subsection, showing that a well-defined Hamiltonian formulation with non-trivial asymptotic symmetries can be found.

The rest of the section is devoted to the massless case, which presents subtle complications.
We will start the discussion of this second case by deriving the fall-off and (strict) parity conditions of the fields and their momenta, which are going to provide a theory with a finite symplectic form, a finite and functionally-differentiable Hamiltonian, and a symplectic action of the Poincar\'e group.
However, these conditions are a bit too strong, in the sense that they do not allow for non-trivial asymptotic symmetries.
We will attempt  to relax the strict parity conditions and discuss which issues arise during the process, that make either the asymptotic symmetry group trivial or the Lorentz boost non-canonical.
Finally, we will make the connection between these issues at spatial infinity and some problems concerning the Lorenz gauge fixing encountered in analyses at null infinity.

\subsection{Massive case}

Let us begin with the derivation of the fall-off conditions of the fields.
As it was done for free electrodynamics in~\cite{Henneaux-ED}, for free Yang-Mills in~\cite{Tanzi:2020}, and for a free complex scalar field in section~\ref{sec:free-scalar}, we are going to derive the fall-off conditions by demanding that they are the most general ones preserved by the action of the Poincar\'e group~(\ref{eoms-begin})--(\ref{eoms-end}).

Focusing on the transformation of $\varphi$ and $\Pi$ and proceeding as in section~\ref{subsec:massless-massive-scalar}, one can show that the massive scalar field needs to vanish at infinity faster than any power-like function, as it happens in the free case.
It is easy to verify, at this point, that the fall-off conditions of $A$ and $\pi$ are exactly those of free electrodynamics~\cite{Henneaux-ED}, that is
\begin{equation}
\begin{aligned} \label{fall-off-free-ED}
  A_r (r,\barr{x}) &= \frac{1}{r} \barr{A}_r (\barr{x}) +\bigo(1/r^2) \,,
  & \pi^r (r,\barr{x}) &= \barr{\pi}^r (\barr{x}) +\bigo(1/r)\,, \\
  A_{\bar{a}} (r,\barr{x}) &=\barr{A}_{\bar{a}} (\barr{x}) +\bigo(1/r) \,, 
  & \pi^{\bar{a}} (r,\barr{x}) &= \frac{1}{r} \barr{\pi}^{\bar{a}} (\barr{x}) + \bigo(1/r^2) \,,
\end{aligned}
\end{equation}
where the results are expressed in radial-angular coordinates.
In addition, the gauge parameter is required to fall off as
\begin{equation} \label{fall-off-gauge}
 \zeta (x) = \barr{\zeta} (\barr{x}) + \bigo (1/r) \,,
\end{equation}
so that the gauge transformations~(\ref{gauge-infinitesimal}) preserve the fall-off conditions of the canonical fields.
Note that this last expression, together with the definition~(\ref{A-perp}) the known fall-off of $N$ and $\vect{N}$ for the Poincar\'e transformations, implies the fall-off
\begin{equation}
    A_\perp (r, \barr{x}) = \frac{1}{r} \barr{A}_\perp (\barr{x}) + \bigo(1/r^2) 
\end{equation}
for the Lagrange multiplier.

Since the scalar field and its momentum quickly vanish at infinity, the asymptotic structure of the theory is effectively the same as in the free electrodynamics case.
This means that proceeding as in~\cite{Henneaux-ED}, one would find a well-defined Hamiltonian formulation of massive-scalar electrodynamics with a canonical action of the Poincar\'e group, and with non-trivial asymptotic symmetries, corresponding to an extension of the Poincar\'e group by the angle-dependent-$\U (1)$ transformations at infinity.

\subsection{Massless case: fall-off and parity conditions} \label{subsec:massless-scalar-ed-fall-off-parity}

As in the massive case, we  begin with the derivation of the fall-off conditions of the fields.
In this case, it is possible to find a power-law ansatz which is preserved by the Poincar\'e transformations~(\ref{eoms-begin})--(\ref{eoms-end}).
Specifically, this corresponds to merging the fall-off conditions of the free massless scalar field~(\ref{fall-off-scalar-massless}) and of free electrodynamics~(\ref{fall-off-free-ED}).
Also in this case, the gauge parameter is required to fall-off as in~(\ref{fall-off-gauge}), so that the gauge transformations~(\ref{gauge-infinitesimal}) preserve the fall-off conditions of the fields.
The asymptotic Poincar\'e transformations of the fields are then found to be
\begin{align}
 %%%% A_r
 \label{poincare-asymptotic-begin}
 \delta_{\xi,\zeta} \barr{A}_r ={}&
 \frac{b\, \barr{\pi}^r}{\sqrt{\barr{\gamma}}}
 + Y^{\bar m} \partial_{\bar m} \barr{A}_r
 \,,\\
 %%%% A_a
 \delta_{\xi,\zeta} \barr{A}_{\bar a} ={}&
 \frac{b \, \barr{\pi}_{\bar a}}{\sqrt{\barr{\gamma}}}
 +Y^{\bar m} \partial_{\bar m} \barr{A}_{\bar a} + \partial_{\bar a} Y^{\bar m} \barr{A}_{\bar m}
 -\partial_{\bar a} \barr{\zeta}
 \,, \\
 %%%% \pi_r
 \delta_{\xi,\zeta} \barr{\pi}^r ={}&
 \barr{\nabla}^{\bar m} \big( b\, \sqrt{\barr{\gamma}}\,
 \partial_{\bar m} \barr{A}_r \big)
 -2 b \sqrt{\barr{\gamma}} \, |\barr{\varphi}|^2 \barr{A}_r 
 +\partial_{\bar m} (Y^{\bar m} \barr{\pi}^r) \,, \\
 %%%% \pi_a
 \label{poincare-asymptotic-pi-a}
 \delta_{\xi,\zeta} \barr{\pi}^{\bar a} ={}& 
 \partial_{\bar m} \big( b \, \sqrt{\barr{\gamma}}\, \barr{F}^{\bar m \bar a})
 -2 b \sqrt{\barr{\gamma}} \, \imaginarypart \left( \barr{\varphi}^* \barr{D}^{\bar a} \barr{\varphi} \right)
 %-2 b \sqrt{\barr{\gamma}} \, |\barr{\varphi}|^2 \barr{A}^{\bar a} 
 +\partial_{\bar m} (Y^{\bar m}\, \barr{\pi}^{\bar a})
 -\partial_{\bar m} Y^{\bar a} \, \barr{\pi}^{\bar m} \,, \\
 %%%% \varphi
 \delta_{\xi,\zeta} \barr{\varphi} ={}& \frac{b \, \barr{\Pi}}{\sqrt{\barr{\gamma}}}
 + Y^{\bar m} \partial_{\bar m} \barr{\varphi}
 + i \barr{\zeta} \, \barr{\varphi} \,, \\
 %%%% \Pi
 \label{poincare-asymptotic-end}
 \delta_{\xi,\zeta} \barr{\Pi} ={}&
 - b\sqrt{\barr{\gamma}} \left( 1 + \barr{A}_r^2 \right) \barr{\varphi}
 + \barr{D}^{\bar m} \left( b\sqrt{\barr{\gamma}} \, \barr{D}_{\bar m} \barr{\varphi} \right)
 %+ i \sqrt{\barr{\gamma}} \left[ b \barr{A}^{\bar a} \partial_{\bar a} \barr{\varphi}
 %+ \barr{\nabla}^{\bar a} \left( b \barr{A}_{\bar a} \barr{\varphi} \right) \right]
 -2 b \sqrt{\barr{\gamma}} \, \lambda |\barr{\varphi}|^2 \, \barr{\varphi}
 + \partial_{\bar m} \left( Y^{\bar m} \barr{\Pi} \right)
 + i \barr{\zeta} \, \barr{\Pi}\,,
\end{align}
where $\barr{\nabla}$ is the covariant derivative of the round unit two sphere and $\barr{D}_{\bar m} \eqdef \barr{\nabla}_{\bar m} + i \barr{A}_{\bar m}$.
%Note that, in the above transformations, the self-interaction $\lambda$ disappears, so that the asymptotic transformations are effectively those of the free fields with the additional minimal couple between the scalar and the electromagnetic field.

The fall-off conditions are not enough to provide a finite symplectic form and a symplectic action of the Poincar\'e group.
In particular, the symplectic form~(\ref{symplectic-form}) still contains two logarithmically-divergent contributions: The first is due to the fall-off conditions of $A$ and $\pi$, while the second is due to the fall-off conditions of $\varphi$ and $\Pi$.
One possible solution to this issue is quite simple.
One merely requires that the asymptotic part of the fields have one definite parity (either even or odd) as functions on the two-sphere at infinity and, then, imposes the opposite parity on their conjugated momenta.
This way, the potentially logarithmically-divergent contributions to the symplectic form are actually zero.
We will see that the presence of the massless scalar field will cause the parity conditions to be slightly more involved.

To fully determine the exact form of the parity conditions, let us remind that they should be such that, not only do they make the symplectic form finite, but also the Poincar\'e transformations symplectic.
Specifically, this happens when $\liephase_X \Omega = 0$, being $\liephase_X$ the Lie derivative in phase space with respect to the vector field $X$ defining the Poincar\'e transformations~(\ref{eoms-begin})--(\ref{eoms-end}).
Using Cartan magic formula and the fact that the symplectic form is closed, one gets
\begin{equation} \label{poincare-canonical-final}
 \liephase_X \Omega =
 \extder (i_X \Omega)= \extder \oint d^2 \barr{x} \;  \sqrt{\barr{\gamma}} \,
 \barr{A}_{r} \left[
 \extder \barr{\nabla}^{\bar m} \left( b \, \barr{A}_{\bar m} \right)
 + 2 b\, \text{Im} \left( \barr{\varphi}^* \extder \barr{\varphi} \right)
 \right] \,,
\end{equation}
after having simplified the expression.
Note that the first summand in the right-hand side of the above expression is precisely the term already appearing in free electrodynamics~\cite{Henneaux-ED}, while the second summand appears due to the presence of the massless scalar field.
We wish to impose parity conditions that make the above expression to vanish identically.
To this end, let us decompose the complex scalar field as
\begin{equation}
 \varphi (x) = \frac{1}{\sqrt{2}} \, \rho (x) e^{i \vartheta (x)} \,.
\end{equation}
The newly-introduced absolute value and the phase of the scalar field need to satisfy the fall-off conditions
\begin{equation} 
 \rho (x) = \frac{1}{r} \barr{\rho} (\barr{x}) + \bigo(1/r^2)
 \qquad \text{and} \qquad
 \vartheta (x) = \barr{\vartheta}  (\barr{x}) + \bigo (1/r) \,,
\end{equation}
in order to be consistent with~(\ref{fall-off-scalar-massless}).
Rewriting~(\ref{poincare-canonical-final}) in terms of these new fields, we see that the Poincar\'e transformations are canonical if
\begin{equation} \label{poincare-canonical-phase}
 \liephase_X \Omega =
 \extder \oint d^2 \barr{x} \;  \sqrt{\barr{\gamma}} \,
 \barr{A}_{r} \left[
 \extder \barr{\nabla}^{\bar m} \left( b \, \barr{A}_{\bar m} \right)
 + b\, \rho^2 \extder \vartheta
 \right]
\end{equation}
vanishes.
This can be achieved in the following way.
First, we require the parity conditions
\begin{equation}
 \barr{A}_r = \barr{A}_r^{\,\text{odd}}
 \qquad \text{and} \qquad
 \barr{\pi}^r = \barr{\pi}^r_{\text{even}} \,,
\end{equation}
so that the related part in the symplectic form is finite and Coulomb is included in the allowed fields configurations.
Note that this choice of parity for $\barr{\pi}^r$ implies that gauge transformations are proper if $\barr{\zeta}$ is an odd function on the sphere and are improper if it is an even function, applying the results of section~\ref{subsec:gauge-transformations}.
Second, one makes~(\ref{poincare-canonical-phase}) to be finite by requiring that
\begin{equation}
 \barr{A}_{\bar m} = \barr{A}_{\bar m}^{\,\text{even}} \,,
 \qquad
 \barr{\pi}^{\bar m} = \barr{\pi}^{\bar m}_{\text{odd}} \,,
 \qquad
 \barr{\vartheta} = \barr{\vartheta}^{\,\text{odd}} \,,
\end{equation}
and that $\barr{\rho}$ is of definite parity, either even or odd.
Note that the parity conditions of $\barr{\vartheta}$ excludes the  improper gauge transformations, such as the constant $U(1)$ at infinity, as these would shift $\barr{\vartheta}$ by an even function.
Finally, in order to make the symplectic form finite, we decompose also the momentum $\Pi$ as
\begin{equation}
 \Pi (x) = \frac{1}{\sqrt{2}} \, R (x) e^{i \Theta (x)} \,,
\end{equation}
which needs to satisfy the fall-off conditions
\begin{equation}
 R (x) = \frac{1}{r} \barr{R} (\barr{x}) + \bigo(1/r^2)
 \qquad \text{and} \qquad
 \Theta (x) = \barr{\Theta}  (\barr{x}) + \bigo (1/r) \,.
\end{equation}
In terms of the absolute values and the phases, the logarithmically-divergent contribution to the symplectic form is
\begin{equation}
    \int \frac{dr}{r} \int d^2 \barr{x} \,
    \Big[ 
    \cos (\barr{\vartheta} - \barr{\Theta}) \Big(\extder \barr{R} \wedge \extder \barr{\rho} + \barr{\rho} \barr{R} \, \extder \barr{\vartheta} \wedge \extder  \barr{\Theta} \Big)
    -\sin (\barr{\vartheta} - \barr{\Theta}) \Big(\barr{\rho} \, \extder \barr{R} \wedge \extder \barr{\vartheta} + \barr{R} \, \extder \barr{\rho} \wedge \extder \barr{\Theta} \Big)
    \Big] \,,
\end{equation}
which vanishes identically once we require that $\barr{R}$  has the opposite parity of $\barr{\rho}$ and that $\barr{\Theta}$ is odd.
Note that also the parity of $\barr{\Theta}$, other than that of $\barr{\vartheta}$, is such that improper gauge transformations are not allowed.
Indeed, in order to preserve these parity conditions, we need to restrict the gauge parameters such that $\barr{\zeta}$ is an odd function.
In turn, this implies that the generator~(\ref{gauge-generator}) is finite and differentiable without the need of a surface term.

Finally, note that the parity conditions that we have just found are preserved by the Poincar\'e transformations.
To see this, one only need to use the asymptotic form of the transformations~(\ref{poincare-asymptotic-begin})--(\ref{poincare-asymptotic-end}) and the equations
\begin{equation}
    \delta \barr{\vartheta} =  \imaginarypart \left( \frac{\delta \barr{\varphi}}{\barr{\varphi}} \right) \,, \qquad
    \delta \barr{\rho} = \barr{\rho} \, \realpart \left( \frac{\delta \barr{\varphi}}{\barr{\varphi}} \right) \,, \qquad
    \delta \barr{\Theta} =  \imaginarypart \left( \frac{\delta \barr{\Pi}}{\barr{\Pi}} \right) \,, \qquad
    \delta \barr{R} =  \barr{R} \, \realpart \left( \frac{\delta \barr{\Pi}}{\barr{\Pi}} \right) \,.
\end{equation}

To sum up, we have seen that, in the massless case, the fields satisfy power-like fall-off conditions.
In order to have a finite symplectic form and a canonical action of the Poincar\'e group, the fall-off conditions need to be complemented with some parity conditions.
We have shown that it is possible to find (strict) parity conditions leading to a well-defined Hamiltonian formulation.
Specifically, the strict parity conditions of $A$ and $\pi$ are the same as those in free electrodynamics~\cite[Sec. 5]{Tanzi:2020}.
The parity conditions of the complex scalar field and its momentum have been found after decomposing them into an absolute value and a phase.
The absolute values of $\varphi$ and $\Pi$ are required to have opposite parity, while the phases need to be both of odd parity.
Notably, the parity conditions imposed on the phases, as well as those on $\barr{A}_{\bar a},$ exclude the improper gauge transformations from the theory and reduces the asymptotic symmetry group to the Poincar\'e group.
In the next subsection, we will try to solve this problem by relaxing the parity conditions.

\subsection{Relaxing the parity conditions}

The solution to reintroduce the possibility of performing improper gauge transformations is quite simple.
Specifically, since the improper gauge transformations are excluded due to the (strict) parity conditions, we simply need relax them so that they are satisfied up to an improper gauge transformations.
Therefore, we require the asymptotic part of the fields that transform non-trivially under gauge transformations to be such that
\begin{equation} \label{parity-conditions-loosen-ED}
 \barr{A}_{\bar{a}} = \barr{A}_{\bar{a}}^{\text{even}}
 -\partial_{\bar a} \barr{\Phi}^{\text{even}} \,,  \qquad
 \barr{\vartheta} = \barr{\vartheta}^{\text{odd}} + \barr{\Phi}^{\text{even}} \,,
 \qquad \text{and} \qquad
 \barr{\Theta} = \barr{\Theta}^{\text{odd}} + \barr{\Phi}^{\text{even}} \,,
\end{equation}
where $\barr{\Phi}^{\text{even}}(\barr{x})$ is an even function on the sphere.
At the same time, the other fields are required to satisfy the same parity conditions as before, that is
\begin{equation}
 \barr{A}_r = \barr{A}_r^{\text{odd}} \,, \qquad
 \barr{\pi}^r = \barr{\pi}^r_{\text{even}} \,, \qquad
 \barr{\pi}^{\bar{a}} =  \barr{\pi}^{\bar{a}}_{\text{odd}} \,,
\end{equation}
while $\barr{R}$ and $\barr{\rho}$ are of definite, and opposite, parity.

These relaxed parity conditions allow for certain the possibility of performing improper gauge transformations, thus extending the asymptotic symmetry group.
However, they also reintroduce back in the theory two issues.
First, the symplectic form is not finite any more.
Indeed, it contains now the logarithmically divergent contribution
\begin{equation}
    \Omega=
    \int \frac{dr}{r}\oint_{S^2} d^2 \barr{x} \;
    \extder \left[ \partial_{\bar a} \barr{\pi}^{\bar a}
    -2 \, \imaginarypart \left( \barr{\Pi}^* \barr{\varphi} \right ) \right]
    \wedge \extder  \barr{\Phi}^{\text{even}}
    + (\text{finite terms})\,.
\end{equation}
To solve this issue, we need merely to note that the term in square brackets in the expression above is nothing else than the leading contribution in the asymptotic expansion of the Gauss constraint~(\ref{gauss-constraint}).
Indeed, it is easy to verify that
\begin{equation}
    \mathscr{G} =
    \frac{1}{r} \left[ \partial_{\bar a} \barr{\pi}^{\bar a}
    -2 \, \imaginarypart \left( \barr{\Pi}^* \barr{\varphi} \right ) \right]
    + \bigo(1/r^2) \defeq
    \frac{1}{r} \barr{\mathscr{G}} + \bigo(1/r^2) \,.
\end{equation}
As a consequence, the symplectic form can be made finite by restricting the phase space to those fields configurations satisfying the further condition $\barr{\mathscr{G}} = 0$.
This does not exclude any solution to the equations of motion, since they already need to satisfy the full Gauss constraint $\mathscr{G} \approx 0$.

The second issue reintroduced after relaxing the parity condition is that the Poincar\'e transformations are not canonical any more.
This is due to the fact that
\begin{equation} \label{poincare-canonical-scalar-ed}
 \liephase_X \Omega =
 \extder \oint d^2 \barr{x} \;  \sqrt{\barr{\gamma}} \,
 \barr{A}_{r} \left[
 \extder \barr{\nabla}^{\bar m} \left( b \, \barr{A}_{\bar m} \right)
 % + b\, \rho^2 \extder \vartheta
 +2 b\, \text{Im} \left( \barr{\varphi}^* \extder \barr{\varphi} \right)
 \right]
\end{equation}
does not identically vanish any more.
In the expression above $\liephase_X \Omega$ is the Lie derivative (in phase space) of the symplectic form $\Omega$ with respect to the vector field $X$, which identifies the Poincar\'e transformations.
In the case of \emph{free} electrodynamics, it was shown that it is possible to make the Poincar\'e transformations canonical once again, by introducing a new boundary degree of freedom $\barr{\Psi}$ and complementing the symplectic form with a boundary term $\omega$~\cite{Henneaux-ED}.
Specifically, this works as follows.
First, one requires that $\barr{\Psi}$ transform under the Poincar\'e transformations as
$\delta_X \barr{\Psi} = \barr{\nabla}^{\bar m} \left( b \, \barr{A}_{\bar m} \right) + Y^{\bar m} \partial_{\bar m} \barr{\Psi}$ and chooses the boundary term to be
\begin{equation} \label{symplectic-boundary}
    \omega = \oint d^2 \barr{x} \, \sqrt{\barr{\gamma}} \, \extder \barr{\Psi} \wedge \extder \barr{A}_r \,,
\end{equation}
so that $\liephase_X (\Omega + \omega) = 0$.
Second, one extends the new field $\barr{\Psi}$ in the bulk and makes $\Psi_{\text{bulk}}$ pure gauge.
The details can be found in~\cite{Henneaux-ED}, where this method was presented for the first time.
Here, we are interested in pointing out that a similar attempt in this case would not be as successful.
Indeed, on the one hand, one would still be able to compensate the first summand in square brackets of~(\ref{poincare-canonical-scalar-ed}).
On the other hand, one would not be able to compensate also the second summand in square brackets, as this is not an exact form.\footnote{
To see precisely that the form is not exact one could either rely on the decomposition into phase and absolute value as done in~(\ref{poincare-canonical-phase}) or on the decomposition into $\varphi_{1,2}$.
We will come back to this point in section~\ref{subsec:Lorentz-Lorenz-SED}.
}

While studying a similar issue in Yang-Mills, we have shown that it is in general not-easily possible to circumvent this type of problems~\cite{Tanzi:2020}.
In that case, we used a general ansatz with quite a few free parameters for the boundary degrees of freedom, for the boundary term of the symplectic form, and for the Poincar\'e transformations of the boundary degrees of freedom and showed that no choice of free parameters was yielding a solution.
In this paper, we will not pursue a similar tedious path.
Rather, we will point out a possible connection between obstructions to a canonical Lorentz boost and some issues in the Lorenz gauge fixing when a flux of charge-current at null infinity is present~\cite{Wald-Satishchandran}, as in the case of a charged massless scalar field with the weakest possible fall-off conditions compatible with the Poincar\'e transformations. 
To this end, we will first analyse some aspects of the free electrodynamics case and, then, deal with the scalar electrodynamics one.

\subsection{The Lorentz boost and the Lorenz gauge: free electrodynamics}
In this subsection, we focus on free electrodynamics and highlights the relation between canonical Poincar\'e transformations and the Lorenz gauge fixing (at infinity).
Since the issues, which were mentioned in the previous subsection, arise because of the boost contribution to the orthogonal part of the Poincar\'e transformations, we will set, after a little introduction, $\vect{N} = 0$ and $N = \xi^\perp = r b (\barr{x}) + T$ in this and in the next subsection.
We will once again start our analysis from the action, but use the knowledge that we have gained so far in the discussion of fall-off and parity conditions.

Let us begin with the action in Lagrangian formulation, which is given by~(\ref{action-Lagrangian}) when setting the scalar field $\varphi$ to zero.
Note that, in principle, the action in the bulk can be complemented by a boundary term.
Let us write it as
\begin{equation}
    (\text{boundary term}) =
    \int dt \oint d^2 \barr{x} \, \mathcal{B} (\barr{x}) \,.
\end{equation}
Note that the above expression implies that we are adding a boundary term at spatial infinity, as it is more fitted for the ensuing discussion, although different types of boundary terms can be and have been considered.
The function $\mathcal{B} (\barr{x})$ depends on (the asymptotic part of) the fields and of $N$.

The variation of the bulk action, in general, will produce other boundary terms due to the necessity of performing some integration by parts while deriving the equations of motion.
In order to have a well-defined action principle, we need to require that, not only do the bulk part of the variation  vanishes producing the bulk equations of motion, but also that the boundary term (which contains also the contribution due to $\mathcal{B}$) of the variation is zero.
One way to deal with the boundary term in the variation of the action is to make it vanish identically by imposing some suitable (fall-off and parity) conditions on the asymptotic behaviour of the fields.
Another way is to make sure that, even if it is not identically zero, it produces boundary equations of motion that do not contain any new information with respect to the bulk ones.
If neither one of the two said situations happens, we end up with some non-trivial equations of motion at the boundary, which could affect the physics of the theory, for instance by trivialising some symmetry.

Before we actually show explicitly the situation in electrodynamics, let us stress that whether or not boundary terms are produced during the variation of the action in the bulk depends on the asymptotic behaviour of the fields, of the lapse $N$, and of the shift $\vect{N}$.
If we had been interested only in the time evolution of the theory, it would have sufficed to consider the fall-off conditions of $N$ and $\vect{N}$ to be as strong as in~\cite[Sec. 2]{RT}.
However, since we wish to have a well-defined action of the Poincar\'e group as well, we need to allow the lapse and the shift to behave asymptotically as $N = \xi^\perp = r b (\barr{x}) + T$ and $N_i = g_{ij} \xi^j$, respectively.
As a consequence, we need to make sure that the variation of the action is well-defined also for these lapse and shift.
Note that, in the following expressions, a dot above a quantity represents the rate of change (with respect to the parameter $t$ of the Hamiltonian flow) of that quantity under the Poincar\'e transformation generated 
by the flow. We are considering the cases 
$N= r b (\barr{x}) +T$ and $\vect{N} = 0$. 
For $b=0$ and $T=1$, this coincides with the 
standard time rate of change. 

Explicitly, the variation of the action is
\begin{equation}
\begin{aligned}
    \extder S ={}&
    \int dt \left\{
    \int d^3 x \left[
    \frac{\sqrt{g}}{N} g^{ab} F_{0b} \, \extder \dot{A}_a
    + \partial_b \left( N \sqrt{g} F^{ba} \right) \extder A_a
    + N \partial_a \left( \frac{\sqrt{g}}{N} g^{ab} F_{0b} \right) \extder A_\perp
    \right] +
    \right. \\
    & \left.+ \oint_{S^2_\infty} d^2 \barr{x} \, \Big[
    - \sqrt{g} \, g^{rm} F_{0m} \, \extder A_\perp
    + N \sqrt{g} F^{a r} \, \extder A_a
    + \extder \mathcal{B}
    \Big]
    \right\} \,,
\end{aligned}
\end{equation}
where the surface integral on $S^2_\infty$ has to be understood as a surface integral over a sphere of radius $R$ followed by the limit $R \rightarrow \infty$.
In the above expression, we have replaced $A_0$ with $A_\perp$ using~(\ref{A-perp}) and we have already performed the needed integration by parts.
The bulk part of the variation lead to the usual equations of motion and symplectic form, which we have already discussed.
Thus, let us focus on the boundary part.
Inserting the usual fall-off conditions of the field and $N = rb +T$, the boundary part of the variation becomes
\begin{equation} \label{variation-boundary}
    \int dt \oint d^2 \barr{x} \, \Big[
    -\sqrt{\barr{\gamma}} \, \dot{\barr{A}}_r \, \extder \barr{A}_\perp 
    + b \sqrt{\barr{\gamma}} \, \partial_{\bar m}  \barr{A}_r \barr{\gamma}^{\bar m \bar n} \, \extder \barr{A}_{\bar n}
    + \extder \mathcal{B}
    \Big] \,,
\end{equation}
where the limit $R \rightarrow \infty$ has already been taken, so that the remaining surface integral is effectively on a unit two-sphere. 
Let us assume for the moment that $\mathcal{B} = 0$, i.e., the action is the usual action of Maxwell electrodynamics without any boundary terms.

On the one hand, we could try to make the expression in~(\ref{variation-boundary}) identically zero by imposing parity conditions on the fields similarly as in section~\ref{subsec:massless-scalar-ed-fall-off-parity}, but this choice would exclude the improper gauge transformations from the theory trivialising the asymptotic symmetry group.
On the other hand, in the absence of parity conditions, the above expression would produce the boundary equations of motion
\begin{equation} \label{boundary-eoms-bad}
    \dot{\barr{A}}_r = 0
    \qquad \text{and} \qquad
    \partial_{\bar m} \barr{A}_r = 0
\end{equation}
so that the variation of the action is well defined for any value of $b$.
The first one of the above equations implies that the asymptotic part of the electric field vanish and, as a consequence, the charges.
One way to see this is to expand the expression in~(\ref{momenta}) to get $\barr{\pi}^r = \dot{\barr{A}}_r \sqrt{\barr{\gamma}} / b = 0$.
In addition, the second equation implies that $\barr{A}_r$ is constant on the sphere.
Since it is also required to be an odd function, in order to have a finite symplectic form, we must conclude that $\barr{A}_r = 0$.
Thus, also in this case, we end up with trivial asymptotic symmetries.

This shows that, if we wish to have a well-defined action principle for Maxwell electrodynamics with the Poincar\'e transformations and non-trivial asymptotic symmetries, we must include a boundary term in the original action.
A suitable choice is
\begin{equation} \label{boundary}
    \mathcal{B} =
    \sqrt{\barr{\gamma}} \, \dot{\barr{A}}_r  \barr{A}_\perp 
    - b \sqrt{\barr{\gamma}} \, \partial_{\bar m}  \barr{A}_r \barr{\gamma}^{\bar m \bar n} \barr{A}_{\bar n}
    %+ b \, \barr{A}_r^2
    \,.
\end{equation}
This boundary term is chosen because it moves all the variations in~(\ref{variation-boundary}) to $\barr{A}_r$ and $\dot{\barr{A}}_r$, so that we obtain one, single boundary equations of motion rather than the two, very-restrictive ones of~(\ref{boundary-eoms-bad}).
A similar boundary term was already considered by Henneaux and Troessaert in~\cite[App. B]{Henneaux-ED}, in order to solve the same issue.
At this point, the boundary part of the variation of the action becomes
\begin{equation} \label{variation-boundary-new}
    \int dt \oint d^2 \barr{x} \, \sqrt{\barr{\gamma}} \,
    %\left\{
    %\barr{A}_\perp \, \extder \dot{\barr{A}}_r 
    %+ \left[ 2 b\, \barr{A}_r + 
    %\barr{\nabla}^{\bar m} \left( b\, \barr{A}_{\bar m} \right) 
    %\right] \extder \barr{A}_r
    %\right\}
    \left[
    \barr{A}_\perp \, \extder \dot{\barr{A}}_r 
    + \barr{\nabla}^{\bar m} \left( b\, \barr{A}_{\bar m} \right) 
    \extder \barr{A}_r
    \right] 
    \,,
\end{equation}
Two things can be noted.
First, when going from the Lagrangian to the Hamiltonian picture as in section~\ref{sec:Lagrangian-Hamiltonian}, $\barr{A}_r$ has now a conjugated momentum on the boundary, namely $\sqrt{\barr{\gamma}} \, \barr{A}_\perp$.
This means that the usual bulk symplectic form needs to be complemented with the boundary term
\begin{equation} \label{symplectic-boundary-new}
    \omega = \oint d^2 \barr{x} \, \sqrt{\barr{\gamma}} \, \extder \barr{A}_\perp \wedge \extder \barr{A}_r \,,
\end{equation}
which coincides with the boundary term~(\ref{symplectic-boundary}) used by Henneaux and Troessaert in~\cite{Henneaux-ED}, after identifying $\barr{A}_\perp$ with $\Psi$.
Note that the need of such a boundary term was first pointed out by Campiglia and Eyheralde in~\cite[Sec. 4]{Campiglia-U1}.

Second, the boundary equation of motion ensuing from~(\ref{variation-boundary-new}) is
$\dot{\barr{A}}_\perp^{\text{odd}} = 
 \barr{\nabla}^{\bar m} \left( b \barr{A}_{\bar m}^{\text{odd}} \right)$, where the equation is restricted to the odd component because $\barr{A}_r$ is odd.
This is nothing else than the odd part of the leading term in the asymptotic expansion of the Lorenz gauge condition ${}^{4} \nabla^\mu A_\mu = 0$, where ${}^4 \nabla$ is the Levi-Civita connection of the four-metric ${}^4 g$ which,
in local coordinates, may be written in the 
form~(\ref{4-metric}) whenever $N\ne 0$.\footnote{Note that at points where $N=0$ 
it is the representation in terms of embedding coordinates that becomes singular, not the 
Lorentz metric as a geometric object.}
To see this let us first compute
\begin{equation}
    \begin{aligned}
        {}^{4} \nabla^\mu A_\mu ={}&
        {}^{4} \nabla_\mu \left( {}^4 g^{\mu \nu} A_\nu \right) 
        = \frac{1}{\sqrt{|{}^4 g|}} \, \partial_\mu \left( \sqrt{|{}^4 g|} \; {}^4 g^{\mu \nu} A_\nu \right) = \\
        ={}& \frac{1}{N \sqrt{g}} \left\{
        \partial_0 \left[ N \sqrt{g} \left( - \frac{1}{N^2}  \right) A_0 \right]
        + \partial_m \left( N \sqrt{g} \, g^{mn} A_m \right)
        \right\} = \\
        ={}& \frac{1}{N} \left[ -\dot A_\perp
        + \nabla^m (N A_m) \right] \,,
    \end{aligned}
\end{equation}
where $\nabla$ is the Levi-Civita connection of the three-dimensional metric $g$.\footnote{
In deriving the expression, we have used twice, once for the four metric ${}^4 g$ and once for the three metric $g$, the fact that, if $\gamma$ is a non-singular metric of any signature on a manifold of any dimension and if $V$ is a vector field, then
\[
 \nabla_m V^m =
 \frac{1}{\sqrt{|\gamma|}} \, \partial_m \left(  \sqrt{|\gamma|} \, V^m \right) \,,
\]
where $\nabla$ is the Levi-Civita connection of $\gamma$.
}
The expansion in powers of $r$ of the expression in square brackets on the last line is then
\begin{equation}
    \frac{1}{r} \left[
    - \dot{\barr{A}}_\perp + 2 b \, \barr{A}_r +
    \barr{\nabla}^{\bar m} \left( b \, \barr{A}_{\bar m} \right) 
    \right] + \bigo(1/r^2)
\end{equation}
and the odd component of this expression, when set to zero, returns the  equation of motion discussed above.
The steps which we used to reach this result are valid in all the regions where $N = r b(\barr{x}) +T$ is not zero.
In terms of the sphere at infinity $S^2_\infty \subset \Sigma$, the condition $N = 0$ identifies a circle orthogonal to the direction of the boost.
Excluding this circle, the image of the sphere at infinity under the one-parameter family of embeddings corresponding to the Lorentz boost sweeps a portion of the hyperboloid at infinity $\mathcal{H}_\infty \subset M$.
Thus, we see that the asymptotic Lorenz condition is imposed on a portion of $\mathcal{H}_\infty$ as a boundary equation of motion, if we wish the boost $b$ to follow from an action principle (or, equivalently, to be canonical).
Extending this requirement to every boost makes the asymptotic Lorenz condition to be imposed on the entire $\mathcal{H}_\infty$.
This show the equivalence of our analysis with that of~\cite[App. B]{Henneaux-ED}.

It is possible, although not strictly necessary, to extend the boundary equation of motion into the bulk (and to the asymptotic even component as well).
To do so, one can proceed as in~\cite[Sec. 4-5]{Kuchar-Stone} and introduce a new contribution to the action
\begin{equation}
    \tilde S [A,\psi] = \int d^4 x \sqrt{- {}^4 g } \; \partial_\alpha \psi \, g^{\alpha \beta} \, A_\beta \,,
\end{equation}
where $\psi$ is a new scalar field.\footnote{
In the analysis done in~\cite{Kuchar-Stone}, the spatial slices of the spacetime are closed manifolds, i.e., compact and without boundary.
Nevertheless, the results of the paper can be applied to our situation as well and are correct up to boundary terms.
}
The variation of the action with respect to $\psi$ yields the desired Lorenz gauge condition in the bulk.
When passing to the Hamiltonian picture, one finds that the conjugated momentum of $\psi$ is $\pi_\psi = - \sqrt{g} A_\perp$, that is the Lagrange multiplier $A_\perp$ up to the density weight.
In order to make sure that the equations of motion in the bulk are not physically affected in the procedure, one needs merely to impose the constraint $\psi \approx 0$.
Without redoing all the computations, let us simply note that, in this way, we obtain exactly the same solution proposed in~\cite{Henneaux-ED}, after identifying $\Psi = A_\perp = - \pi_\psi /\sqrt{g}$ and $\pi_\Psi = \sqrt{g} \, \psi \approx 0$.
Note that the constraint $\pi_\Psi \approx 0$ induces gauge transformations that shift the value of $\Psi = A_\perp$ by an arbitrary function in the bulk, so that the bulk part of the Lorenz condition ${}^{4} \nabla^\mu A_\mu = 0$ can be violated arbitrarily.
However, on the boundary this is not the case since shifting $\barr{\Psi}$ by an odd function is not a gauge transformation, but rather a true symmetry of the theory.
Finally, note that this procedure introduces two new canonical degrees of freedom: the orthogonal component of the vector potential $A_\perp$, which has been elevated from being a mere Lagrange multiplier to a true degree of freedom, and a momentum conjugated to it.\footnote{
Due to the constraint $\pi_\Psi$ and the gauge symmetry ensuing from it, however, the only physically-relevant degree of freedom that has be introduced is the odd component of $\barr{A}_\perp$.
}

To sum up, we have shown that the action of Maxwell electrodynamics needs to be complemented by a boundary term, if one wishes to have a well-defined action principle, which works also with the lapse and shift given by the Poincar\'e transformations, featuring non-trivial asymptotic symmetries.
A suitable choice for the boundary term is~(\ref{boundary}), which, once added to the original action, leads to two consequences.
First, when deriving the symplectic form, one finds that it contains the boundary term~(\ref{symplectic-boundary-new}) as in~\cite{Henneaux-ED}.
Second, one gets a new, non-trivial boundary equation of motion, which is nothing else than the leading term in the asymptotic expansion of the Lorenz gauge condition.
In addition, changing this gauge fixing at infinity by shifting $\barr{\Psi}$ by an odd function is not a proper gauge transformation, but rather a true symmetry of the theory, as thoroughly explained in~\cite{Henneaux-ED}.
Let us now focus on the case in which a charged massless scalar field is present.

\subsection{The Lorentz boost and the Lorenz gauge: scalar electrodynamics} \label{subsec:Lorentz-Lorenz-SED}

Let us now reintroduce the massless scalar field minimally-coupled to electrodynamics.
Proceeding as in the previous subsection, we consider the variation of the action~(\ref{action-Lagrangian}) and split it into a bulk part and a boundary part.
The part in the bulk provides the equations of motion  in the bulk, after some integration by parts.
The boundary part of the variation, at this point, reads
\begin{equation} \label{variation-boundary-scalar-ed}
\begin{aligned}
    \int dt \oint d^2 \barr{x} \, \Big[&
    -\sqrt{\barr{\gamma}} \, \dot{\barr{A}}_r \, \extder \barr{A}_\perp 
    + b \sqrt{\barr{\gamma}} \, \partial_{\bar m}  \barr{A}_r \barr{\gamma}^{\bar m \bar n} \, \extder \barr{A}_{\bar n} + \\
    &+2 b \sqrt{\barr{\gamma}} \, \realpart \left( \barr{\varphi}^* \extder \barr{\varphi} \right)
    -2 b \sqrt{\barr{\gamma}} \, \barr{A}_r \imaginarypart \left( \barr{\varphi}^* \extder \barr{\varphi} \right)
    + \extder \mathcal{B}
    \Big] \,,
\end{aligned}
\end{equation}
where, again, we are allowing the presence of a boundary term $\mathcal{B}$ in the action.
The first line of the expression above contains the contribution due to free electrodynamics, which was amply discussed in the previous subsection.
The second line appears due to the presence of the scalar field and is made up of two contributions.
The former does not bring any issue, as it can be readily absorbed into $\extder \mathcal{B}$.
Indeed,
\begin{equation}
    2 \realpart \left( \barr{\varphi}^* \extder \barr{\varphi} \right) = 
    \barr{\varphi}_1 \extder \barr{\varphi}_1 +
    \barr{\varphi}_2 \extder \barr{\varphi}_2 =
    \extder \left( \frac{\barr{\varphi}_1^2 + \barr{\varphi}_2^2}{2} \right)
    = \extder \left( \barr{\varphi}^* \, \barr{\varphi} \right)\,.
\end{equation}
The second term is the one causing all the troubles.
Indeed, not only cannot it be rewritten as a total variation, but also it cannot be written as $\barr{A}_r$ times the total variation of something, since
\begin{equation}
    2 \imaginarypart \left( \barr{\varphi}^* \extder \barr{\varphi} \right) = 
    \barr{\varphi}_1 \extder \barr{\varphi}_2 -
    \barr{\varphi}_2 \extder \barr{\varphi}_1 \,,
\end{equation}
which is not even a closed one form.
If we had been able to rewrite the term in the variation as $\barr{A}_r \extder \mathcal{B}'$, we could have included a term $- \barr{A}_r \mathcal{B}'$ into $\mathcal{B}$ and we would have obtain a single, more relaxed equation of motion at the boundary, as in the previous subsection.
This equation would have been the Lorenz gauge condition at infinity modified by a contribution coming from $\mathcal{B}'$.
When passing to the Hamiltonian formalism, the above issue translate in the fact that the Lorentz boost fails to be canonical due to the presence of a boundary term in $\liephase_X \Omega$, unless strict parity conditions are impose, \textit{de facto} trivialising the asymptotic algebra.

We propose a connection between the impossibility of having a canonical Lorentz boost when asymptotic symmetries are allowed, i.e. when we impose the relaxed parity conditions, and some issues related to the Lorenz gauge fixing when a flux of charge-current at null infinity is present~\cite{Wald-Satishchandran}.
Although we will not provide a formal proof of this statements, we will provide two indications that this is the case.

Before we present the two arguments, let us summarise the relevant results described by Wald and Satishchandran in~\cite{Wald-Satishchandran}.
Specifically, they have analysed the case of electrodynamics in four and higher and shown that, due to the fall-off conditions of the fields, it is not possible to find a Lorenz gauge fixing --- i.e. it does not exist a gauge parameter satisfying the correct fall-off conditions and bringing the four potential in Lorenz gauge --- if the dimension of the spacetime is four and a flux of charge-current at null infinity is present.
This setup is expected in our situation, due to the presence of a charged massless scalar field satisfying the most general fall-off at spatial infinity, compatible with the Poincar\'e transformations.
Note that no obstruction to the Lorenz gauge fixing is present in higher dimension, even in the presence of a charge-flux at null infinity.

The first argument which we provide is that, in the case of free electrodynamics, the Lorenz gauge fixing at infinity appears as a boundary equations of motion that needs to be imposed if we wish, at the same time, a well-defined action principle (also for the lapse and shift of the Poincar\'e transformations) and non-trivial asymptotic symmetries.
The same equation cannot be derived if a massless scalar field is present, as we have seen in the first part of this subsection.

The second argument is that no issue arises in the variation of the action (or, equivalently, in the Lorentz boost being canonical) in higher dimensions.
In this paper, we have worked exclusively in $3+1$ dimensions, as this is the physically-relevant case.
However, it is possible to repeat the same analyses in higher dimensions, as in the case of free electrodynamics, which was already studied by Henneaux and Troessaert in~\cite{Henneaux-ED-higher}.
So, let us assume for the remainder of this subsection that the spacetime dimension is $n+1$, being $n$ an odd number.

We can derive also in this case the fall-off conditions of the fields by requiring that they are power-like and that they are preserved by the Poincar\'e transformations, obtaining\footnote{
We work in radial angular components $(r, \barr{x})$, where $\barr{x}$ are coordinates on the unit $(n-1)$-sphere.
}
\begin{equation} \label{fall-off-higher-dim}
\begin{aligned}
  A_r (r,\barr{x}) &= \frac{1}{r^{n-2}} \barr{A}_r (\barr{x}) +\bigo \left( 1/r^{n-1} \right) \,,
  & \pi^r (r,\barr{x}) &= \barr{\pi}^r (\barr{x}) +\bigo(1/r)\,, \\
  A_{\bar{a}} (r,\barr{x}) &= \partial_{\bar a} \barr{\Phi} (\barr{x}) +
  \frac{1}{r^{n-3}} \barr{A}_{\bar{a}} (\barr{x}) +\bigo \left( 1/r^{n-2} \right) \,, 
  & \pi^{\bar{a}} (r,\barr{x}) &= \frac{1}{r} \barr{\pi}^{\bar{a}} (\barr{x}) + \bigo \left( 1/r^2 \right) \,, \\
  \varphi (r, \barr{x}) &= \frac{1}{r^{(n-1)/2}} \barr{\varphi} + \bigo \left( 1/r^{(n+1)/2} \right) \,,
  & \Pi (r, \barr{x}) &= \frac{1}{r^{(3-n)/2}} \barr{\Pi} + \bigo\left( 1/r^{(5-n)/2} \right)\,,
\end{aligned}
\end{equation}
The first two lines of the above expression contain precisely the findings of~\cite{Henneaux-ED-higher}.
Two things can be noted about them.
First, if $n > 3$, the fall-off conditions are enough to ensure that the symplectic form is finite (see~\cite{Henneaux-ED-higher} for a detailed discussion), so that no parity condition is needed.
Second, $A_{\bar a}$ contains two relevant asymptotic parts: a zeroth-order contribution, which is a gradient of a function on the $(n-1)$-sphere, and a contribution of order $1/r^{n-3}$.
If $n = 3$, as it is in the previous part of this section, the gradient can be reabsorbed in $\barr{A}_{\bar a}$, but this is not possible if $n > 3$.
Finally, the last line of the expression above contains the fall-off conditions of the scalar field and its momentum.
These lead to a logarithmic divergence in the symplectic form which can be dealt with by means of parity conditions.

Ignoring the details about these subtleties, let us show directly that the scalar field does not bring any obstruction to a canonical Lorentz boost.
To this end, let us compute the Lie derivative of the symplectic form with respect to the vector field of the Poincar\'e transformations.
After a few passages, we find
\begin{equation} \label{canonical-poincare-higher-dim}
    \liephase_X \Omega =
    \oint_{S^{n-1}_\infty} d^{n-1} \barr{x} \left[
    \xi^\perp \sqrt{g} \, \extder F^{ra} \wedge \extder A_a
    + 2 \sqrt{g} \, \xi^\perp \, \realpart \left(
    \extder D_r \varphi \wedge \extder \varphi^*
    \right)
    \right] \,,
\end{equation}
where the integration over $S^{n-1}_\infty$ has to be understood as an integration over an $(n-1)$-sphere of radius $R$ followed by the limit $R \rightarrow \infty$.
Also in the case of higher dimensions, we see that the Poincar\'e transformations fail to be canonical due to a boundary contribution.

The first term in square brackets in equation~(\ref{canonical-poincare-higher-dim}) is the contribution due to free electrodynamics.
Using the fall-off conditions~(\ref{fall-off-higher-dim}), it reduces to
\begin{equation}
     \oint d^{n-1} \barr{x} \left\{
    - b \sqrt{\barr{\gamma}} \, \barr{\gamma}^{\bar m \bar n} \, \extder
    \left[ (n-3) \barr{A}_{\bar m} + \partial_{\bar m} \barr{A}_r \right]
    \wedge \extder \partial_{\bar n} \barr{\Phi}
    \right\} \,,
\end{equation}
where the integration is now performed on a unit $(n-1)$-sphere, as the limit $R \rightarrow \infty$ has been already taken.
This contributions has been already thoroughly analysed in~\cite{Henneaux-ED-higher} and, basically, can be dealt with by introducing a new boundary degree of freedom, similarly to the case of free electrodynamics in four dimension, which we have discussed in the previous subsection.
The second term, on the contrary, is the new contribution due to the massless scalar field.
Expanding it with the use of the fall-off conditions~(\ref{fall-off-higher-dim}), it reduces to
\begin{equation}
     \lim_{R \rightarrow \infty} \oint_{S^{n-1}_R} d^{n-1} \barr{x} \left\{
    - \frac{1}{R^{n-3}} \, 2 b \sqrt{\barr{\gamma}} \; \imaginarypart \left[
    \extder \left( \barr{\varphi} \barr{A}_r \right) \wedge \extder \barr{\varphi}^*
    \right]
    \right\} \,,
\end{equation}
which vanishes if $n > 3$ and produces the problematic term of~(\ref{poincare-canonical-scalar-ed}) if $n = 3$.
Thus, we have shown that no issue is present if $n > 3$ even if there is a massless charged field.

In summary, we have studied the situation of scalar electrodynamics in this section.
We have seen that is the scalar field is massive, the analysis and the asymptotic symmetries do not differ from the free electrodynamics case, which is already well known~\cite{Henneaux-ED}.
A massless scalar field, however, brings some complications.
Specifically, despite it is possible to provide a well-defined Hamiltonian formulation of the theory with canonical Poincar\'e transformations, this does not include any non-trivial asymptotic symmetry, due to the strict parity conditions required.
Relaxing the parity conditions in order to allow improper gauge transformation and keeping the symplectic form finite is possible, but at the cost of making the Lorentz boost non canonical.
We have identified a possible explanation for the failure of having, at the same time, a canonical action of the Poincar\'e group and non-trivial asymptotic symmetries in the impossibility of imposing a Lorenz gauge condition if there is a flux of charge-current at null infinity~\cite{Wald-Satishchandran}. Interestingly, 
this fact is a peculiarity of the physically-relevant
four-dimensional spacetime and does not happen in 
higher dimensions.
We have provided two evidences in support of this hypothesis.
First, the importance of the Lorenz gauge condition at infinity in free electrodynamics with canonical Poincar\'e transformations and non-trivial asymptotic symmetries.
Second, the fact that neither the obstruction to the Lorenz gauge fixing nor the issues in having a canonical action of the Poincar\'e group are present in higher dimensions.
This concludes our analysis of the asymptotic structure of scalar electrodynamics using the Hamiltonian formulation. In the next section we will focus on 
the abelian Higgs model.

\section{Abelian Higgs model} 
\label{sec:abelian-Higgs}

In this section, we wish to study the asymptotic symmetries of the theory described by the Hamiltonian~(\ref{Hamiltonian}), when $\mu^2 > 0$ and $\lambda > 0$.
This choice of the parameters leads to the Mexican-hat potential for  the scalar field and to the abelian Higgs mechanism.
Let us begin by determining the fall-off behaviour of the fields.

\subsection{Fall-off conditions of the fields} \label{subsec:Higgs-fall-off}

Let us begin the discussion about the abelian Higgs model by studying the asymptotic behaviour of the fields and, in particular, their fall-off conditions.
As usual, we wish to find the ``largest'' phase space which is stable under the action of the Poincar\'e transformations.
The derivation of the fall-off conditions is very similar to that presented in section~\ref{subsec:Mexican-hat-potential} and differs only in the last steps and in the fact that one needs to take into consideration a greater number of fields, as we have to include the abelian one-form potential $A_a$ and its conjugated momentum $\pi^a$ in the discussion.
This will have an effect also on the fall-off conditions of the phase of the scalar field, which will turn out to be a bit different from those of section~\ref{subsec:Mexican-hat-potential}.

First of all, let us note that we need the phase space to contain the minimum-energy solutions to the equations of motion, as these are physically relevant solutions.
Specifically, this means that the phase space needs to include at least the solutions
\begin{equation}
 A_a (x) = 0 \,, \qquad
 \pi^a (x) = 0 \,, \qquad
 \varphi (x) = \varphi^{(\vartheta)} (x) \,, \qquad \text{and} \qquad
 \Pi (x) = 0 \,,
\end{equation}
where the constant solution $\varphi^{(\vartheta)} (x) \eqdef v/ \sqrt{2} \exp (i \vartheta)$ was already defined in equation~(\ref{solutions-ssb}).
We already know that one consequence of this fact is that the potential~(\ref{potential}) needs to be corrected by the addition of the constant $\lambda v^4 /4 $, being $v \eqdef \sqrt{ \mu^2 / \lambda }$, so that it becomes
\begin{equation} \label{potential-new}
    V(\varphi^* \varphi) = \lambda \left(\frac{v^2}{2} -\varphi^* \varphi \right)^2 \,.
\end{equation}
Another consequence is that we have to exclude the trivial solution to the equation of motion --- i.e. all fields and momenta equal to zero --- from phase space, for otherwise the Hamiltonian would not be finite.

Secondly, the fall-off conditions are expressed more effectively when the $\varphi(x)$ is expressed in terms of its absolute value and phase.
So, let us write
\begin{equation} \label{scalar-abs-phase}
 \varphi (x) = \frac{1}{\sqrt{2}} \rho (x) \, e^{i \vartheta (x)}
\end{equation}
At this point, we only need to proceed in the same way as in section~(\ref{subsec:Mexican-hat-potential}) excluding the last step, in which the behaviour of $\vartheta (x)$ was determined.
With the same arguments, we conclude also in this case that $\rho (x) = v + h (x)$, where $h (x)$ is quickly vanishing up to the second derivative order, and that $\Pi (x)$ is quickly vanishing.

Thirdly, let us determine the fall-off behaviour of the phase $\vartheta (x)$.
As in section~\ref{subsec:Mexican-hat-potential}, let us consider the transformation of $\Pi$ under time evolution, i.e., equation~(\ref{eoms-end}) at $N=1$ and $\vect{N}=0$.
Up to terms that are quickly vanishing at infinity, we find
\begin{equation}
\begin{split}
    \delta \Pi = \varphi &\left\{
    - \sqrt{g} \, g^{ab} ( \partial_a \vartheta + A_a ) (\partial_b \vartheta + A_b)
    +i D_a \left[ \sqrt{g} \, g^{ab} (\partial_b \vartheta + A_b) \right]
    \right\}\\
    + &(\text{quickly-vanishing terms}) \,.
\end{split}
\end{equation}
The above transformation preserves the fall-off condition of $\Pi$ so long as $\partial_a \vartheta + A_a$ is quickly vanishing together with its first-order derivatives.
So, let us write $A = - d \vartheta + \tilde A$, where $\vartheta$ is only required to have a well-defined limit $\barr{\vartheta} (\barr{x}) = \lim_{r \rightarrow \infty} \vartheta (x) $ as a function on the sphere at infinity, whereas $\tilde A$ is a quickly-vanishing function together with its first derivatives.
Note that the Lagrange multiplier $N A_\perp$ needs to satisfy the same fall-off conditions of $\vartheta$.

Lastly, we need to determine the fall-off behaviour 
of $\pi^a$.
To do so, one merely need to demand that the fall-off behaviour of $A= -d \vartheta +\tilde A$ is preserved by a generic Lorentz boost.
One sees that the only possibility is to require that $\pi^a$ is quickly vanishing.
In turn, this fall-off behaviour is preserved by the Poincar\'e transformations so long as the second derivatives of $\tilde A$, too, are quickly vanishing.
This concludes the discussion about the fall-off conditions of the fields in the abelian Higgs model.

To sum up, we have shown that, if one splits the scalar field into an absolute value and a phase as in~(\ref{scalar-abs-phase}), the former has to be $\rho (x) = v + h (x)$, where $h(x)$ is quickly vanishing up to its second-order derivatives.
The phase $\vartheta (x)$, on the contrary is merely required to have a well-defined limit
$\barr{\vartheta} (\barr{x}) = \lim_{r \rightarrow \infty} \vartheta (x)$
as a function on the sphere at infinity and the same holds true for the Lagrange multiplier $N A_\perp$.
In addition, the one-form $A$ can be written as $A= -d \vartheta + \tilde A$, where $\tilde A$ is quickly vanishing up to its second-order derivatives.
Finally, the momenta $\pi^a$ need to be quickly vanishing.
In particular, note that the fall-off behaviour of the one-form $A_a$ and of its momentum $\pi^a$ is substantially different from that of electrodynamics, either in the free case~\cite{Henneaux-ED,Tanzi:2020} or when coupled to a scalar field (see section~\ref{sec:scalar-electrodynamics}).
These fall-off conditions ensure that the Poincar\'e transformations have a well defined action on the phase space.

\subsection{Well-defined Hamiltonian formulation and symmetries} \label{subsec:Higgs-Hamiltonian-symmetries}

Having derived the fall-off conditions of the fields, we can now provide the well-defined Hamiltonian formulation of the abelian Higgs model.
In particular, we will provide the exact form of the Hamiltonian, of the generator of the Poincar\'e transformations, and of the generator of the gauge transformations.
Furthermore, we will also identify the asymptotic symmetries of the theory.

To begin with, let us note that the symplectic form~(\ref{symplectic-form}) is finite, thanks to the quick fall-off of the fields.
For the same reason,  both the Hamiltonian and the generator of the Poincar\'e transformations are finite and differentiable.
These can be inferred from the generator
\begin{equation} \label{Hamiltonian-Higgs}
H[A,\pi,\varphi,\Pi;g,N,\vect{N};A_\perp]=
 \int  d^3 x \Big[ 
 N \mathscr{H}
 + N^i \mathscr{H}_i
 \Big] \,,
\end{equation}
by setting $N=1$, $\vect{N} = 0$ (Hamiltonian) and $N=\xi^\perp$, $N^i = \xi^i$ (Poincar\'e generator).
In the above generator
\begin{equation} 
\begin{aligned} \label{superHamiltonian-Higgs}
 \mathscr{H} \eqdef& 
 \frac{\pi^a \pi_a + \Pi_1^2 + \Pi_2^2}{2\sqrt{g}}
 + \frac{\sqrt{g}}{4} F_{ab} F^{ab}
 -A_\perp \, \mathscr{G}
 + \frac{\sqrt{g}}{2} g^{ab} \big( \partial_a \varphi_1 \partial_b \varphi_1 + \partial_a \varphi_2 \partial_b \varphi_2 \big) + \\
 & + \sqrt{g} A^a \big( \varphi_1 \partial_a \varphi_2 - \varphi_2 \partial_a \varphi_1 \big)
 +\frac{1}{2} A_a A^a \big( \varphi_1^2 + \varphi_2^2 \big)
 + \sqrt{g} \, V(\varphi^* \varphi)
\end{aligned}
\end{equation}
is responsible for the orthogonal transformations and
\begin{equation} 
 \mathscr{H}_i \eqdef
 \pi^a\partial_i A_a - \partial_a (\pi^a A_i)
 + \Pi_1 \partial_i \varphi_1 + \Pi_2 \partial_i \varphi_2
\end{equation}
is responsible for the tangential transformations.
Note that the potential is 
\begin{equation}
    V(\varphi^* \varphi) = \lambda \left( \frac{v^2}{2} - \varphi^* \varphi \right)^2 \,,
\end{equation}
which differs from the original potential~(\ref{potential}) due to the addition of the constant $\lambda v^4 /4$, so that the energy of the vacuum solutions to the equations of motion is finite.
% \begin{equation} 
%  H[A,\pi;g,\zeta]=
%  \int d^3 x \Big[ 
%  \mathscr{H}
%  +\zeta \, \mathscr{G}
%  \Big]
% \end{equation}
% and the generator of the Poincar\'e transformations
% \begin{equation} 
%  H[A,\pi;\xi,g,\zeta]=
%  \int d^3 x \Big[ 
%  \xi^\perp \mathscr{H} + \xi^i  \mathscr{H}_i
%  +\zeta \, \mathscr{G}
%  \Big]
% \end{equation}
% are finite and differentiable with respect to the canonical fields.
% In the above expressions,
% \begin{equation} 
% \begin{aligned}
%  \mathscr{H} \eqdef& 
%  \frac{\pi^a \pi_a}{2\sqrt{g}}
%  +\frac{\Pi^* \Pi}{\sqrt{g}}
%  + \frac{\sqrt{g}}{4} F_{ab} F^{ab}
%  + \sqrt{g} g^{ab} \partial_a \varphi^* \partial_b \varphi +\\
%  & - \sqrt{g} A^a \big( \varphi^* \partial_a \varphi -  \varphi \partial_a \varphi^* \big)
%  + 2 A_a A^a \varphi^* \varphi
%  + \sqrt{g} \, V(\varphi^* \varphi)
% \end{aligned}
% \end{equation}
% contributes to the orthogonal part of the transformations,
% \begin{equation} 
% \begin{aligned}
%  \mathscr{H}_i \eqdef& 
%  \pi^a\partial_i A_a - \partial_a (\pi^a A_i) + \Pi^* \partial_i \varphi + \Pi \partial_i \varphi^*
% \end{aligned}
% \end{equation}
% contributes to the tangential transformations, and
% \begin{equation}
%  \mathscr{G} \eqdef \partial_a \pi^a + \varphi^* \Pi - \varphi \Pi^* \weq 0
% \end{equation}
% is the Gauss constraint, contributing to the gauge transformations.

Furthermore, let us note that the Gauss constraint $\mathscr{G} (x)$ --- which has the same expression as in~(\ref{gauss-constraint}) --- appears in the generator~(\ref{superHamiltonian-Higgs}) multiplied by the Lagrange multiplier $A_\perp$.
In general, gauge transformations are generated by
\begin{equation} \label{generator-gauge-Higgs}
 G[\zeta] \eqdef \int d^3 x \,  \zeta (x) \mathscr{G} (x) \weq 0 \,,
\end{equation}
which is finite and differentiable without the need of any surface term.
Note that, in the above generator, $\zeta$ is only required to have a well-defined limit $\barr{\zeta} (\barr{x}) = \lim_{r \rightarrow \infty} \zeta (x)$, so that the transformations~(\ref{gauge-infinitesimal}) preserve the fall-off conditions identified in the previous subsection.

Two things can be noted at this point.
First, the phase $\vartheta$ can always be trivialised by a proper gauge transformation, so that it carries no physical meaning.
Specifically, from~(\ref{gauge-finite}), we see that $\Gamma_{-\vartheta} (\varphi) = \rho /\sqrt{2}$, without any phase.\footnote{
Note that this is a complete gauge fixing.
}
Second, since~(\ref{generator-gauge-Higgs}) is already finite and differentiable without the need of any boundary term, it cannot be extended to a generator of improper gauge transformations, contrary to the case of electrodynamics~\cite{Henneaux-ED}.
As a consequence, the asymptotic symmetries of the theories are trivially the Poincar\'e transformations.
Indeed, the only generator of asymptotic symmetries is $H [\xi, \vect{\xi}]$ which satisfies the algebra
\begin{equation}
%\begin{align} \label{GG-bracket}
%  \big\{ G[\zeta_1] , G[\zeta_2] \big\} &= 0 \,,
%  \\
%  \label{HG-bracket}
%  \big\{ H[\xi^\perp,\vect{\xi}] , G[\zeta] \big\} &= G[\lie_{\vect{\xi}}\zeta] \,, \\
 \label{HH-bracket}
 \big\{ H[\xi^\perp_1,\vect{\xi}_1] , H[\xi^\perp_2,\vect{\xi}_2] \big\} %&
 = H[\hat \xi^\perp,\hat{\vect{\xi}}] + G [\hat \zeta] \,,
%\end{align}
\end{equation}
where
\begin{equation}
 \hat \xi^\perp = \lie_{\vect{\xi}_1} \xi^\perp_2 - \lie_{\vect{\xi}_2} \xi^\perp_1 \,,
 \qquad
 \hat \xi^m = \tilde \xi^m + [\vect{\xi}_1,\vect{\xi}_2]^m \,,
 \qquad
 \hat \zeta = A_m \tilde \xi^m
 + \xi_1 \lie_{\vect{\xi_2}} A_\perp - \xi_2 \lie_{\vect{\xi_1}} A_\perp \,.
\end{equation}
Here,
$\tilde \xi^i \eqdef g^{ij} (\xi^\perp_1 \partial_j \xi^\perp_2 - \xi^\perp_2 \partial_j \xi^\perp_1)$
(which simplifies the expressions above and the following discussion), $\lie$ is the Lie derivative on spatial slices, and $[\vect{\xi}_1,\vect{\xi}_2]$ is the commutator of the vector fields $\vect{\xi}_1$ and $\vect{\xi}_2$.
The above algebra is easily seen to be a Poisson-representation of the Poincar\'e algebra up to (proper) gauge transformations, due to the presence of the constraint on the right-hand side of~(\ref{HH-bracket}).
The fact that the Poincar\'e algebra is recovered up to proper gauge transformations is not in general a problem (see e.g. the discussion in~\cite[Sec. 2]{Beig}).

Before we conclude this section, let us note that the $\hat \zeta$ in the expressions above depends on the canonical fields and, in particular, on $A_m$.\footnote{We remind that $A_\perp$ is not a canonical field, but only a Lagrange multiplier.}
As a consequence the transformation generated by $G[\hat \zeta]$ slightly differs from the usual gauge transformations.
Specifically, it induces the transformations
\begin{equation}
 \delta A_a = - \partial_a \hat \zeta
 \,,
 \qquad
 \delta \pi^a = - \tilde \xi^a \, \mathscr{G} \approx 0  \,,
 \qquad
 \delta \varphi =  i \hat \zeta \, \varphi  \,,
 \qquad \text{and} \qquad
 \delta \Pi = i \hat \zeta \, \Pi \,.
\end{equation}
It is useful to compare the above transformations with those caused by $\zeta$ in equations~(\ref{gauge-infinitesimal}).
Two things emerge.
First, $A$, $\varphi$, and $\Pi$ transform in the same way, with the only difference being that the parameter $\hat \zeta$ is field-dependent.
Secondly, the transformation of $\pi$ due to $\hat \zeta$ is not trivial any more.
Nevertheless, it is proportional to the Gauss constraint and, thus, vanishes on the constraint hypersurface.
Note that the transformations above deserve by all means the title of gauge transformations, as one part of them is generated by the constraints
\begin{equation}
 A_m \mathscr{G} \weq 0
\end{equation}
smeared with $\tilde \xi^m$, while the other part of them is generated by the usual Gauss constraint $\mathscr{G}$ smeared by $\xi_1 \lie_{\vect{\xi_2}} A_\perp - \xi_2 \lie_{\vect{\xi_1}} A_\perp$.
Let us neglect this second part, as it is of a well-known shape, and focus on the first one, whose generator
$\tilde G[\tilde{\vect{\xi}}] \eqdef G[\tilde \xi^m A_m]$ is easily seen to be well defined and functionally differentiable with respect to the canonical fields, as the fields are rapidly vanishing while approaching spatial infinity.
Furthermore, it satisfies the algebra
\begin{equation}
 \big\{ \tilde G[\tilde{\vect{\xi}}_1] , \tilde G[\tilde{\vect{\xi}}_2] \big\} = \tilde G[\tilde{\vect{\xi}}] \,,
 \qquad \text{where} \qquad
 \tilde{\vect{\xi}} = [\tilde{\vect{\xi}}_1 , \tilde{\vect{\xi}}_2] \,.
\end{equation}

This concludes the discussion about the asymptotic symmetries of the abelian Higgs model.
To sum up, we have shown that the fall-off conditions derived in the previous subsection are enough to ensure a well-defined Hamiltonian formulation with a canonical action of the Poincar\'e group.
Moreover, we have seen that the phase $\vartheta$ can always be trivialised by a proper gauge transformation and that the asymptotic symmetries of the abelian Higgs model are trivial, in the sense that the asymptotic-symmetry group is the Poincar\'e group.
This was shown by computing the Poisson-algebra of $H[\xi,\vect{\xi}]$, which is a Poisson representation of the Poincar\'e algebra up to proper gauge transformations.
Before we draw our conclusions, let us briefly comment on the fate of the Goldstone boson, which emerged as a consequence of the spontaneous symmetry breaking of the global $\U (1)$ in section~\ref{subsec:Mexican-hat-potential}.

\subsection{The fate of the Goldstone boson}

At the end of section~\ref{subsec:Mexican-hat-potential}, we discussed that, in the case of the spontaneous symmetry breaking of the global $\U (1)$ symmetry, the action could be rewritten in terms of two real scalar fields: the massive $h$ and the massless $\chi$.
The former was identified to be the candidate Higgs field in the abelian Higgs model, while the latter was recognised as the Goldstone boson.

Let us repeat that analysis for the abelian Higgs model using the fall-off conditions of section~\ref{subsec:Higgs-fall-off}.
Proceeding as in section~\ref{subsec:Mexican-hat-potential}, let us consider the action in the Lagrangian picture, which can be obtained from~(\ref{action-Lagrangian}) by adding the constant $\lambda v^4 /4$ to the potential.
Rewriting this action in terms of $h$, $\vartheta$, and $\tilde A$, we obtain
\begin{equation} \label{action-expansion-Higgs}
\begin{aligned}
    S[h,\vartheta,\tilde A] ={}& \int d^4 x \left\{
    - \frac{1}{2} \left( {}^4 g^{\alpha \beta} \partial_{\alpha} h \, \partial_\beta h + 2 \mu^2 h^2 \right) + \right. \\
    &- \left. \left( \frac{1}{4} {}^4 g^{\alpha \gamma} \, {}^4 g^{\beta \delta}  \tilde{F}_{\alpha \beta} \tilde{F}_{\gamma \delta}
    + \frac{v^2}{2} \, {}^4 g^{\alpha \beta} \tilde{A}_{\alpha} \tilde{A}_{\beta} \right)
    +(\text{interactions})
    \right\} \,,
\end{aligned}
\end{equation}
where the interactions include all the terms that are not quadratic in the fields.
In the expression above, we have introduced $\tilde{A}_0 \eqdef A_0 +\dot \vartheta$, whose quickly-falling asymptotic behaviour can be inferred from that of the momentum $\Pi$, and $\tilde{F} \eqdef d \tilde{A}$.

Three things can be noted from the expression above.
First, there is a real scalar fields $h$ of squared mass $m_h^2 \eqdef 2 \mu^2$, which corresponds to the Higgs field.
Second, the spin-one field $\tilde{A}$ becomes massive with a squared mass $m_A^2 \eqdef v^2$.
The mass $m_A$ of the spin-one field depends on the vacuum expectation value $v$ of the complex scalar field $\varphi$ and, in general, on the coupling of the Higgs to the original gauge potential $A$ (in this paper, it was set to the value of $1$).
Last, but not least, there is no trace of a massless scalar field, which could play the role of the Goldstone boson.

The disappearance of the Goldstone boson can be tracked down precisely to the choice $A = - d \vartheta + \tilde A$, which we did in section~\ref{subsec:Higgs-fall-off}.
On the one hand, this choice makes the fall-off condition of the momentum $\Pi$ to be preserved by the Poincar\'e transformations.
On the other hand, it makes the gauge-covariant derivative of $\varphi$ to be independent of $\vartheta$, so that the action~(\ref{action-expansion-Higgs}) is also independent of the phase $\vartheta$.\footnote{We remind that in section~\ref{subsec:Mexican-hat-potential}, the role of the Goldstone boson was played by the part $\chi$ of the phase $\vartheta$ which was quickly falling at infinity.
}
Therefore, if we wished to reintroduce the Goldstone boson, we would have to modify slightly the fall-off conditions of section~\ref{subsec:Higgs-fall-off}.

To this end, let us write the phase $\vartheta = \vartheta' + \chi/v$ as the sum of two parts.
The former of the two, $\vartheta'$, is the ``power-like'' part of $\vartheta$, while the latter, $\chi /v$, is the ``quickly-falling'' part.
The only requirement while performing this split is that $\chi /v$ is actually a quickly-falling function.
Two things can be noted.
First, the fall-off behaviour of $\Pi$ is preserved by the Poincar\'e transformations so long as $A = -d \vartheta' +\tilde A$, being $\tilde A$ quickly falling.
When this choice is introduced in the action~(\ref{action-Lagrangian}), we get the expression
\begin{equation} \label{action-expansion-Higgs-Goldstone}
\begin{aligned}
    S[h,\chi,\tilde A] ={}& \int d^4 x \left\{
    - \frac{1}{2} \left( {}^4 g^{\alpha \beta} \partial_{\alpha} h \, \partial_\beta h + 2 \mu^2 h^2 \right)
    - \frac{1}{2} {}^4 g^{\alpha \beta} \partial_{\alpha} \chi \, \partial_\beta \chi +  \right. \\
    & \left. - \left( \frac{1}{4} {}^4 g^{\alpha \gamma} \, {}^4 g^{\beta \delta}  \tilde{F}_{\alpha \beta} \tilde{F}_{\gamma \delta}
    + \frac{v^2}{2} \, {}^4 g^{\alpha \beta} \tilde{A}_{\alpha} \tilde{A}_{\beta} \right) 
    +(\text{interactions})
    \right\} \,,
\end{aligned}
\end{equation}
rather than~(\ref{action-expansion-Higgs}).
The expression above does indeed contain the massless Goldstone boson $\chi$, other than the already-present Higgs field $h$ and massive spin-one field $\tilde A$.
Second, the split of $\vartheta$ into a power-like part $\vartheta'$ and a quickly-falling part $\chi /v$ is obviously ambiguous.
This was not the case in section~\ref{subsec:Mexican-hat-potential}, since, in that case, the only allowed power-like part of $\vartheta$ was its asymptotic value $\barr{\vartheta}$ on the sphere at infinity, which could be unequivocally identified by $\barr{\vartheta} \eqdef \lim_{r \rightarrow \infty} \vartheta$.
The main consequence of this ambiguity in the splitting of $\vartheta$ into $\vartheta'$ and $\chi$ is that $\Omega$ becomes degenerate, so that it is a pre-symplectic form rather than a symplectic one.
Indeed, one can easily check that $i_Y \Omega = 0$, if $Y$ is chosen so that
\begin{equation}
    \delta_Y \vartheta' = \zeta \,, \qquad
    \delta_Y \chi = - v \zeta \,, \qquad
    \delta_Y \tilde A = d \zeta \,, \qquad \text{and} \qquad
    \delta_Y (\text{other fields}) = 0 \,,
\end{equation}
for any quickly-falling $\zeta$.
At this point, one would need to deal with this issue as in~\cite{Henneaux:Rarita-Schwinger}.

In this paper, we have preferred not to pursue this path, since it would introduce some mathematical complications without any advantage on the physical side.
Indeed, as we have seen in section~\ref{subsec:Higgs-Hamiltonian-symmetries}, the phase $\vartheta$ can always be set to zero by means of a proper gauge transformation (with gauge parameter $-\vartheta$).
As a consequence, neither $\vartheta'$ nor $\chi$ are physically-relevant fields.

This concludes the discussion concerning the abelian Higgs model.
To summarise this section, we have derived the fall-off conditions of the fields and shown that these lead to a well-defined Hamiltonian formulation of the theory with a canonical action of the Poincar\'e group.
As a consequence of the quick fall-off behaviour of the fields, the proper gauge transformations cannot be extended to improper ones and the asymptotic symmetry group trivially coincide with the Poincar\'e group.
Furthermore, we have seen that the various fields can be interpreted as a massive spin-one field, a Higgs field, and a Goldstone boson.
The latter, although absent due to the chosen fall-off conditions, can be reintroduced by a slight modification of these.
Nevertheless, it is physically irrelevant, since it can be trivialised by means of a proper gauge transformation.

\section{Conclusions} 
\label{sec:conclusions}
We consider the results of this paper to be 
both, interesting and encouraging. The results 
concerning massive scalar fields were clearly hoped 
for and it is encouraging to see that this hope 
was fulfilled in an unambiguous way, thereby 
providing further confidence into the Hamiltonian 
method for the analysis of asymptotic structures 
and symmetries. As already discussed at length in 
\cite{Tanzi:2020}, the obvious and characteristic 
advantage of this method is to embed the discussion 
on asymptotic symmetries into a formalism of 
clear-cut rules and interpretation. The result for 
the massless case was not a surprise, though we had
no firm intuition whether we should expect it. In that sense, we consider it interesting. 
 
Although the models considered here are not at the 
forefront of physical phenomenology, the abelian 
model does provides good insight into what to expect 
in other Higgs models, such as the physically-relevant 
case of the  electroweak sector. We provided ample
discussion of these expectations.  
In fact, one may speculate that similar results hold 
in the case of the abelian mechanism of the 
electroweak theory, that is 
$\SU(2)_L\times\U(1)_Y \rightarrow \U(1)_{\text{e.m.}}$, where $\SU(2)_L$ is the isospin acting on the left-handed fermions, $U(1)_Y$ is generated by hypercharge, 
and  $\U(1)_{\text{e.m.}}$ by electric charge.
Let $W^I$, $I=1,2,3$, be the standard components of the connection associated to 
$\SU(2)_L$ and $B$ the one associate to $\U(1)_Y$.
Then, one can rewrite
\begin{align}
 A &= \sin \theta_W W^3 + \cos \theta_W B \,, \\
 Z &= \cos \theta_W W^3 - \sin \theta_W  B \,,\\
 W^{\pm} &= \frac{1}{\sqrt{2}} \left( W^1 \mp i W^2 \right) \,,
\end{align}
where $\theta_W$ is the Weinberg angle.
Due to the Higgs mechanism, the equations of motion 
(and the Poincar\'e transformations) of $Z$ and 
$W^\pm$ will contain an effective mass term, while 
no such term will be present in the equations of 
$A$.

Given that, one may expect to find that $A$ has a 
power-like behaviour, while $Z$ and $W^\pm$ are 
quickly vanishing. This expectations is due to the 
fact that the behaviour at infinity seems to depend 
on whether or not a mass term (or an effective mass
term) is  present, and \emph{not} on the specific 
field under consideration. In this paper, this is 
what happens to the free scalar field and to the 
one form $A$ in the abelian Higgs mechanism.
A consequence of the above-mentioned fall-off conditions is that the equations of motion (and the Poincar\'e
transformations) of $A$ and its conjugated momentum 
$\pi$ become those of free electrodynamics near spatial
infinity. Therefore, one may be led to the conjecture 
that the discussion on parity conditions simply 
reduces to that already presented by Henneaux and
Troessaert.

Another way to generalise the field content, that 
would also be of particular interest in view of our
previous results in \cite{Tanzi:2020}, is to consider 
the $\SU(2)$-Yang-Mills-Higgs case. This is currently 
under investigation and will follow soon.

\acknowledgments
% \section*{Acknowledgements}
% \noindent
Support by the DFG Research Training Group 1620 ``Models of Gravity'' is gratefully acknowledged.

\appendix

\section{Fall-off behaviour of massive fields} \label{appendix:massive-fall-off}
We would like to show that, in the massive case, the fall-off behaviour of the field and the momentum needs to be decreasing more rapidly than any power-like function.
Specifically, let us denote with $\mathbb{P}$ the phase space consisting of all the allowed field configurations $(\varphi,\Pi)$ and with $\Poi$ the Poincar\'e group.
In order to have a well-defined relativistic field theory, we need to require that the action of any Poincar\'e transformation maps points belonging to the phase space into points belonging to the phase space.\footnote{
The phase space $\mathbb{P}$ should be though of as 
a sub-manifold in some infinite-dimensional manifold 
of functions which are sufficiently regular so that 
the explicit expressions of the Poincar\'e transformations~(\ref{eoms-complex-begin})--(\ref{eoms-complex-end}) make sense.
}
In other words, we have to impose the condition that, for all $g \in \Poi$, one has  $g \, \mathbb{P} \subseteq \mathbb{P}$.\footnote{Note that we 
require the group $\Poi$ to operate on 
$\mathbb{P}$ by a \emph{group action}, which means 
that the map 
$\Poi\times\mathbb{P}\rightarrow\mathbb{P}$, 
$(g,p)\mapsto gp$, satisfies $g(hp)=(gh)p$ and 
$ep=p$ (where $e$ is the group identity) for 
all $g,h\in\Poi$ and all $p\in\mathbb{P}$. 
This immediately implies that, for 
any $g\in\Poi$, the map $\mathbb{P}\rightarrow\mathbb{P}$, $p\mapsto gp$ is a bijection. Hence $g\mathbb{P}\subseteq\mathbb{P}$ is, in fact, equivalent to 
$g\mathbb{P}=\mathbb{P}$.}
%
%\footnote{
%Note that, the statement $\forall g \in \Poi$, $g\, 
%\mathbb{P} \subseteq \mathbb{P}$ is actually 
%equivalent to $\forall g \in \Poi$, $g\, \mathbb{P} = \mathbb{P}$.
%In order to prove the equivalence of the two
% statements, we only need to show that the former %implies the latter, as the other implication is %obviously true.
%Let us do so.
%Due to the hypothesis, we only need to verify that 
%$\forall g \in \Poi$, $g\, \mathbb{P} \supseteq
% \mathbb{P}$, that is $\forall g \in \Poi$, 
%$\forall p \in \mathbb{P}$, $\exists p' \in 
%\mathbb{P}$  such that $p = g p'$.
%One sees immediately that $p' \eqdef g^{-1} p$ belongs %to $\mathbb{P}$ due to the hypothesis and fulfils the %identity.
%\hfill$\blacksquare$
%}
In this appendix, we wish to show that this requirement, together with the finiteness of the Hamiltonian and the at-most-logarithmic divergence of the symplectic form, implies that,
%%%BEGIN%%%Version 1: mostly written in words%%%
% for all $(\varphi,\Pi) \in \mathbb{P}$ and for all $\alpha,\beta \in \integers$, one has
% \begin{equation} \label{thesis-fall-off}
%  r^\alpha \varphi (x) \rightarrow 0
%  \qquad \text{and} \qquad
%  r^\beta \Pi (x) \rightarrow 0
% \end{equation}
%%%END%%%
%%%BEGIN%%%Version 2: written in symbols. Thesis in the formula%%%
\begin{equation} \label{thesis-fall-off}
 \forall (\varphi,\Pi) \in \mathbb{P} \,, \;
 \forall \alpha,\beta \in \integers \,, \quad
 r^\alpha \, \varphi (x) \rightarrow 0
 \quad \text{and} \quad
 r^\beta \, \Pi (x) \rightarrow 0
\end{equation}
%%%END%%%
in the limit $r \eqdef |x| \rightarrow \infty$.\footnote{
In order to avoid issues in the ensuing proof, we need to assume that the phase space $\mathbb{P}$ is not empty.
This is easily achieved by assuming that $(\varphi^{(0)},\Pi^{(0)}) \in \mathbb{P}$, being $\varphi^{(0)} (x) = 0$ and $\Pi^{(0)} (x) = 0$.
This field configuration, other than being the minimum-energy solution to the equations of motion, is also invariant under the action of the Poincar\'e group and satisfies the statement in~(\ref{thesis-fall-off}).
}

To this end, let us focus only on a part of the full Poincar\'e transformations~(\ref{eoms-complex-begin})--(\ref{eoms-complex-end}) and, specifically on
\begin{equation} 
 \delta' \varphi = \xi^\perp \frac{\Pi}{\sqrt{g}}
 \qquad \text{and} \qquad
 \delta' \Pi = - \xi^\perp \sqrt{g} \, m^2 \varphi \,.
\end{equation}
When considering only a Lorentz boost, i.e. $\xi^\perp = r \, b(\barr{x})$, and writing explicitly the dependence on the radial and angular coordinates, the above expressions become
\begin{equation} \label{boost-special-case}
 \delta' \varphi (r,\barr{x}) =
 \frac{b(\barr{x})}{\sqrt{\barr{\gamma}(\barr{x})}} \, \frac{\Pi(r,\barr{x})}{r} 
 \qquad \text{and} \qquad
 \delta' \Pi (r,\barr{x}) =
 - b (\barr{x}) \sqrt{\barr{\gamma}(\barr{x})} \;m^2 \, r^3 \, \varphi (r,\barr{x}) \,.
\end{equation}

Let us define for all $(\varphi,\Pi) \in \mathbb{P}$ the quantities
\begin{equation}
 \alpha_\varphi \eqdef
 \sup \big\{ \alpha \in \integers \colon r^\alpha \varphi (x) \rightarrow 0 \big\}
 \qquad \text{and} \qquad
 \beta_\Pi \eqdef
 \sup \big\{ \beta \in \integers \colon r^\beta \Pi (x) \rightarrow 0 \big\} \,.
%  \text{ as } r\rightarrow \infty
\end{equation}
First of all, let us note that these quantities are well defined.
Indeed, the finiteness of the mass term in the Hamiltonian (proportional to $\varphi^* \varphi$) implies that $\varphi (x) \rightarrow 0$, whereas the finiteness of the kinetic term (proportional to $\Pi^2$) implies that $r^{-1} \Pi (x) \rightarrow 0$.
Therefore, the sets on the right-hand sides of the definitions above are not empty and the suprema exist.
Note that, with the same argument, we can also conclude that
$\alpha_\varphi \ge 0$ and $\beta_\Pi \ge -1$ for every $(\varphi,\Pi)$ belonging to the phase space.

There are two possibilities for $\alpha_\varphi$ and, analogously, for $\beta_\Pi$.
First, the value of $\alpha_\varphi$ may be $+\infty$, in which case $\varphi$ is according to the statement~(\ref{thesis-fall-off}) that we wish to prove.
Secondly, it may happen that $\alpha_\varphi$ is a finite integer number, in which case the supremum is actually a maximum and $r^{\alpha_\varphi + 1} \varphi$ converges to some function on the sphere.
In principle, this function on the sphere can be divergent, but need not be identically zero.

In order to prove the original statement~(\ref{thesis-fall-off}), we need to show that, for all $(\varphi, \Pi) \in \mathbb{P}$, both $\alpha_\varphi$ and $\beta_\Pi$ are infinite.
To this purpose, let us define
\begin{equation}
 \barr{\alpha} \eqdef \min \{ \alpha_\varphi \colon \varphi \in \mathbb{P}_{|\varphi} \}
 \qquad \text{and} \qquad
 \barr{\beta} \eqdef \min \{ \beta_\Pi \colon \Pi \in \mathbb{P}_{|\Pi}\} \,,
\end{equation}
which are well-defined quantities, since $\alpha_\varphi \ge 0$ and $\beta_\Pi \ge -1$ for every $(\varphi,\Pi) \in \mathbb{P}$, so that the sets on the right-hand sides of the definitions above are non-empty subsets of $\integers \cup \{ +\infty \}$ bounded from below and, as a consequence, the minima exist.
Note that the value of $\barr{\alpha}$ and $\barr{\beta}$ can actually be infinite.
This happens, respectively, when $\alpha_\varphi = +\infty$ for all $\varphi$ and when $\beta_\Pi = +\infty$ for all $\Pi$.
It is easy to see that the statement~(\ref{thesis-fall-off}) is equivalent to the case in which both $\barr{\alpha}$ and $\barr{\beta}$ are infinite.

Let us assume, \emph{ad absurdum}, that at least one among $\barr{\alpha}$ and $\barr{\beta}$ is finite.
To begin with, we note that also the other quantity need to be finite.
This can be seen as follows.
Let us assume that $\barr{\alpha} \in \integers$ and let $(\varphi, \Pi) \in \mathbb{P}$ be such that $\alpha_\varphi = \barr{\alpha}$.\footnote{
The existence of $(\varphi, \Pi) \in \mathbb{P}$ satisfying
$\alpha_\varphi = \barr{\alpha}$  is guaranteed by the fact that $\barr{\alpha}$ is a minimum.
}
After applying a Poincar\'e transformation, we reach the field configuration $(\varphi',\Pi')$ which still belongs to the phase space $\mathbb{P}$ due to the hypothesis.
From the second equation in~(\ref{boost-special-case}), it follows that $\Pi'$ contains the term
\begin{equation}
 \delta' \Pi (r,\barr{x}) =
 - b (\barr{x}) \sqrt{\barr{\gamma}(\barr{x})} \; m^2 \, r^3 \, \varphi (r,\barr{x}) \,,
\end{equation}
which is easily seen to satisfy
$r^{\barr{\alpha} -3} \, \delta' \Pi \rightarrow 0$, while
$r^{\barr{\alpha} -2} \, \delta' \Pi$ does not converge to zero.
Since this is only one of the terms composing $\Pi'$, we cannot make an exact statement about the value of $\beta_{\Pi'}$, but we can nevertheless conclude that $\beta_{\Pi'} \le \barr{\alpha} - 3$, which implies
\begin{equation} \label{inequality-beta}
 \barr{\beta} \le \barr{\alpha} -3 \,,
\end{equation}
showing that $\barr{\beta}$ is finite if $\barr{\alpha}$ is finite.
Analogously, one can show that, if $\barr{\beta}$ is finite, also $\barr{\alpha}$ is finite and satisfies the inequality
\begin{equation} \label{inequality-alpha}
 \barr{\alpha} \le \barr{\beta} + 1 \,.
\end{equation}
The combination of the two inequalities~(\ref{inequality-beta}) and~(\ref{inequality-alpha}) readily yields us the contradiction
\begin{equation}
 \barr{\alpha} \le
 \barr{\beta} + 1 \le
 (\barr{\alpha} - 3) +1 =
 \barr{\alpha}-2 \,.
\end{equation}
Hence, we must conclude that both $\barr{\alpha}$ and $\barr{\beta}$ are infinite, which proves the statement~(\ref{thesis-fall-off}), as we wished.
\hfill$\blacksquare$

\bibliography{biblio}{}

\providecommand{\href}[2]{#2}\begingroup\raggedright\begin{thebibliography}{10}

\bibitem{Tanzi:2020}
Roberto Tanzi and Domenico Giulini, \emph{Asymptotic symmetries of {Yang-Mills}
  fields in {Hamiltonian} formulation},
  \href{https://doi.org/10.1007/JHEP10(2020)094}{\emph{Journal of High Energy
  Physics} {\bfseries 10} (2020) 94}
  [\href{https://arxiv.org/abs/2006.07268}{{\ttfamily arXiv:2006.07268}}].

\bibitem{Henneaux-ED}
Marc Henneaux and C{\'e}dric Troessaert, \emph{Asymptotic symmetries of
  electromagnetism at spatial infinity},
  \href{https://doi.org/10.1007/jhep05(2018)137}{\emph{Journal of High Energy
  Physics} {\bfseries 5} (2018) 137}
  [\href{https://arxiv.org/abs/1803.10194}{{\ttfamily arXiv:1803.10194}}].

\bibitem{Henneaux-ED-higher}
Marc Henneaux and C{\'e}dric Troessaert, \emph{Asymptotic structure of
  electromagnetism in higher spacetime dimensions},
  \href{https://doi.org/10.1103/physrevd.99.125006}{\emph{Physical Review D}
  {\bfseries 99} (2019) 125006}
  [\href{https://arxiv.org/abs/1903.04437}{{\ttfamily arXiv:1903.04437}}].

\bibitem{Wald-Satishchandran}
Gautam Satishchandran and Robert~M. Wald, \emph{Asymptotic behavior of massless
  fields and the memory effect},
  \href{https://doi.org/10.1103/PhysRevD.99.084007}{\emph{Physical Review D}
  {\bfseries 99} (2019) 084007}
  [\href{https://arxiv.org/abs/1901.05942}{{\ttfamily arXiv:1901.05942}}].

\bibitem{Kuchar:1976c}
Karel Kucha\v{r}, \emph{{Kinematics of tensor fields in hyperspace.~III}},
  \href{https://doi.org/https://doi.org/10.1063/1.522978}{\emph{Journal of
  Mathematical Physics} {\bfseries 17} (1976) 801}.

\bibitem{Kuchar:1976a}
Karel Kucha\v{r}, \emph{{Geometry of hyperspace.~I}},
  \href{https://doi.org/https://doi.org/10.1063/1.522976}{\emph{Journal of
  Mathematical Physics} {\bfseries 17} (1976) 777}.

\bibitem{Kuchar:1976b}
Karel Kucha\v{r}, \emph{{Kinematics of tensor fields in hyperspace.~II}},
  \href{https://doi.org/https://doi.org/10.1063/1.522977}{\emph{Journal of
  Mathematical Physics} {\bfseries 17} (1976) 792}.

\bibitem{Giulini:SpringerHandbookSpacetime}
Domenico Giulini, \emph{Dynamical and {Hamiltonian} formulation of {General
  Relativity}},  in \emph{Springer Handbook of Spacetime}, Abhay Ashtekar and
  Vesselin Petkov, eds., (Berlin), pp.~323--362, Springer Verlag, (2014),
  \href{https://doi.org/https://doi.org/10.1007/978-3-642-41992-8_17}{DOI}.

\bibitem{Dirac-book}
Paul Adrien~Maurice Dirac, \emph{Lectures on Quantum Mechanics}. Belfer
  Graduate School of Science, 1964.

\bibitem{Kuchar-Stone}
Karel~V. Kucha\v{r} and Christopher~L. Stone, \emph{A canonical representation
  of spacetime diffeomorphisms for the parametrised {Maxwell} field},
  \href{https://doi.org/10.1088/0264-9381/4/2/013}{\emph{{Classical and Quantum
  Gravity}} {\bfseries 4} (1987) 319}.

\bibitem{Teitelboim-YM2}
Rafael Benguria, Patricio Cordero, and Claudio Teitelboim, \emph{Aspects of the
  {Hamiltonian} dynamics of interacting gravitational gauge and higgs fields
  with applications to spherical symmetry},
  \href{https://doi.org/10.1016/0550-3213(77)90426-6}{\emph{Nuclear Physics B}
  {\bfseries 122} (1977) 61}.

\bibitem{RT}
Tullio Regge and Claudio Teitelboim, \emph{Role of surface integrals in the
  {Hamiltonian} formulation of general relativity},
  \href{https://doi.org/10.1016/0003-4916(74)90404-7}{\emph{Annals of Physics}
  {\bfseries 88} (1974) 286}.

\bibitem{Campiglia-U1}
Miguel Campiglia and Rodrigo Eyheralde, \emph{Asymptotic {U(1)} charges at
  spatial infinity},
  \href{https://doi.org/10.1007/jhep11(2017)168}{\emph{Journal of High Energy
  Physics} {\bfseries 11} (2017) 168}
  [\href{https://arxiv.org/abs/1703.07884}{{\ttfamily arXiv:1703.07884}}].

\bibitem{Beig}
Robert Beig and Niall \'o Murchadha, \emph{The {Poincar\'e} group as the
  symmetry group of canonical general relativity},
  \href{https://doi.org/10.1016/0003-4916(87)90037-6}{\emph{Annals of Physics}
  {\bfseries 174} (1987) 463}.

\bibitem{Henneaux:Rarita-Schwinger}
Oscar Fuentealba, Marc Henneaux, Sucheta Majumdar, Javier Matulich, and Turmoli
  Neogi, \emph{Asymptotic structure of the {Rarita-Schwinger} theory in four
  spacetime dimensions at spatial infinity},
  \href{https://doi.org/https://doi.org/10.1007/JHEP02(2021)031}{\emph{Journal
  of High Energy Physics} {\bfseries 32} (2021)}.

\end{thebibliography}\endgroup
\bibliographystyle{JHEP-fullnames-erratum-arxiv}
% \begin{thebibliography}{00}
% \end{thebibliography}
\end{document}